  \pdfoutput=1
  \documentclass[graphicx,amssymb,amsmath,enumerate,11pt]{report}
  \usepackage{fancyhdr}
  \fancyheadoffset[]{0.1cm}

  \usepackage{epsfig}
  \usepackage{axodraw}
  \usepackage{url}
  \newcommand\mchapter[2]{\chapter*{#1}
  \vskip -0.5cm \noindent {\it \LARGE #2}
  \addcontentsline{toc}{chapter}{#1\\{\normalsize\it #2}}}
\def\ltap{\ \raisebox{-.4ex}{\rlap{$\sim$}} \raisebox{.4ex}{$<$}\ }
\def\gtap{\ \raisebox{-.4ex}{\rlap{$\sim$}} \raisebox{.4ex}{$>$}\ }

\newcommand{\bea}{\begin{equation}\begin{array}{c}}
\newcommand{\eea}{\end{array}\end{equation}}
\newcommand{\ea}{\end{array}} 
\newcommand{\beq}{\begin{equation}}
\newcommand{\eeq}{\end{equation}}
\newcommand{\bad}{\begin{array}{ccc}}
\newcommand{\dmsol}{\mbox{$\Delta m^2_{\odot}$}}
\newcommand{\dma}{\mbox{$\Delta m^2_{\rm atm}$ }}
\newcommand{\mefff}{\mbox{$ < \!~ m ~\! > $}}
\newcommand{\eV}{\mbox{$~{\rm eV}$}}

\newcommand{\betabeta}{\mbox{$(\beta \beta)_{0 \nu}$}}
\newcommand{\meff}{\mbox{$\left|  \langle \,\,  m \,\, \rangle \right| $}}
\usepackage{slashed}
\usepackage{multirow} 

\newcommand{\be}{\begin{equation}}
\newcommand{\ee}{\end{equation}}

\def \gsim{\mathrel{\vcenter
     {\hbox{$>$}\nointerlineskip\hbox{$\sim$}}}}

\def \lesssim{\mathrel{\vcenter
     {\hbox{$<$}\nointerlineskip\hbox{$\sim$}}}}

\def\gappeq{\mathrel{\rlap {\raise.5ex\hbox{$>$}}
{\lower.5ex\hbox{$\sim$}}}}
\def\lappeq{\mathrel{\rlap{\raise.5ex\hbox{$<$}}
{\lower.5ex\hbox{$\sim$}}}}

\begin{document}      

 \rhead{\bfseries The Nature of Massive Neutrinos}

\mchapter{The Nature of Massive Neutrinos}
{Author:\ S. T. Petcov$^{a,b}$}\hspace{-0.05cm}
\footnote{Invited review article for the special issue
on Neutrino Physics of the journal
Advances in High Energy Physics. 
}

\label{ch-01:nature}

\vspace{0.5cm}

\begin{center}
$^a$ {\it SISSA/INFN, Via Bonomea 265, I-34136 Trieste, Italy} \\ [6pt]
$^b$ {\it Kavli IPMU, University of Tokyo, Tokyo, Japan}
\end{center}

\vspace{3cm}

\begin{center}
{\bf Abstract}
\end{center}
The compelling experimental evidences for 
oscillations of solar, reactor, atmospheric 
and accelerator neutrinos imply the 
existence of 3-neutrino mixing in 
the weak charged lepton current.
The current data on the 3-neutrino mixing
parameters are summarised and 
the phenomenology 
of 3-$\nu$ mixing is reviewed.
The properties of massive 
Majorana neutrinos and of their various 
possible couplings are discussed in detail.
Two models of neutrino mass generation with 
massive Majorana neutrinos - the type I 
see-saw and the Higgs triplet model, 
are briefly reviewed.
The problem of determining the nature - 
Dirac or Majorana, of massive neutrinos 
is considered.
The predictions for the effective
Majorana mass $\meff$ in 
neutrinoless double beta
($\betabeta-$) decay
in the case of 3-neutrino mixing 
and massive Majorana neutrinos 
are summarised.
The physics potential of the 
experiments, searching for $\betabeta-$decay 
for providing information on the type of 
the neutrino mass spectrum,
on the absolute scale of neutrino masses 
and on the Majorana CP-violation 
phases in the PMNS  neutrino 
mixing matrix, is also briefly 
discussed. The opened questions
and the main goals of future 
research in the field of neutrino 
physics are outlined.

\newpage

%


\section{Introduction: The Three Neutrino Mixing (an Overview)}
\label{sec-01:intro}

  It is a well-established experimental fact that
the neutrinos and antineutrinos
which take part in the standard
charged current (CC) and neutral
current (NC)  weak interaction
are of three varieties (types) or flavours:
electron, $\nu_e$ and $\bar{\nu}_e$,
muon, $\nu_\mu$ and $\bar{\nu}_\mu$, and
tauon, $\nu_\tau$ and  $\bar{\nu}_\tau$.
The notion of neutrino type or flavour
is dynamical: $\nu_e$ is the neutrino
which is produced with $e^+$,
or produces an $e^-$
in CC weak interaction
processes;
$\nu_\mu$ is the neutrino which
is produced with $\mu^+$, or
produces  $\mu^-$, etc.
The flavour of a given neutrino
is Lorentz invariant.
Among the three different flavour neutrinos and
antineutrinos, no two are identical.
Correspondingly, the states which describe
different flavour neutrinos must be orthogonal
(within the precision of the current data):
$\langle \nu_{l'} |\nu_l\rangle = \delta_{l'l}$,
$\langle \bar{\nu}_{l'} |\bar{\nu}_l\rangle = \delta_{l'l}$,
$\langle \bar{\nu}_{l'} |\nu_l\rangle = 0$.

  It is also well-known from the existing data
(all neutrino experiments were done
so far with relativistic neutrinos
or antineutrinos), that the flavour
neutrinos $\nu_l$ (antineutrinos $\bar{\nu}_l$), are
always produced in weak interaction
processes in a state that is predominantly 
left-handed (LH) (right-handed (RH)).
To account for this fact, $\nu_l$ and $\bar{\nu}_l$
are described in the Standard Model (SM) by a chiral LH flavour
neutrino field $\nu_{lL}(x)$, $l=e,\mu,\tau$.
For massless $\nu_l$, the
state of $\nu_l$ ($\bar{\nu}_l$), which
the  field  $\nu_{lL}(x)$ annihilates (creates),
is with helicity (-1/2) (helicity +1/2).
If $\nu_l$ has a non-zero mass $m(\nu_l)$,
the state of $\nu_l$ ($\bar{\nu}_l$)  is a linear
superposition of the helicity (-1/2) and (+1/2)
states, but the helicity +1/2 state (helicity (-1/2) state)
enters into the superposition with a coefficient $\propto m(\nu_l)/E$, 
E being the neutrino energy, and thus is strongly 
suppressed. Together with the LH 
charged lepton field $l_{L}(x)$, $\nu_{lL}(x)$ forms
an $SU(2)_{L}$ doublet in the Standard 
Model. In the absence of neutrino 
mixing and zero neutrino masses, 
$\nu_{lL}(x)$ and $l_{L}(x)$ can be assigned 
one unit of the additive lepton charge 
$L_l$ and the three charges $L_l$, $l=e,\mu,\tau$, 
are conserved by the weak interaction.

  At present there is no compelling evidence 
for the existence of states of relativistic neutrinos 
(antineutrinos), which are predominantly 
right-handed, $\nu_R$ (left-handed, $\bar{\nu}_L$). 
If RH neutrinos and LH antineutrinos 
exist, their interaction with matter 
should be much weaker than the weak 
interaction of the flavour LH neutrinos 
$\nu_l$ and RH antineutrinos $\bar{\nu}_l$, 
i.e., $\nu_R$ ($\bar{\nu}_L$) should be 
``sterile'' or ``inert'' neutrinos 
(antineutrinos)~\cite{01-BPont67}. 
In the formalism of the Standard 
Model, the sterile $\nu_R$ and $\bar{\nu}_L$ 
can be described by $SU(2)_L$ singlet 
RH neutrino fields $\nu_R(x)$. 
In this case, $\nu_R$ and $\bar{\nu}_L$ 
will have no gauge interactions, i.e., 
will not couple to the weak $W^{\pm}$ 
and $Z^0$ bosons. If present in an 
extension of the Standard Model 
(even in the minimal one), 
the RH neutrinos can play a crucial role 
i) in the generation of neutrino masses and mixing, 
ii) in understanding the remarkable disparity between 
the magnitudes of neutrino masses and 
the masses of the charged leptons and quarks, 
and iii) in the generation of the observed 
matter-antimatter asymmetry of the 
Universe (via the leptogenesis mechanism~\cite{01-LG}, see 
also, e.g., \cite{01-LG1}).  
In this scenario which is based on the 
see-saw theory~\cite{01-seesaw}, there is a link between
the generation 
of neutrino masses and the generation of the 
baryon asymmetry  of the Universe. 
The simplest hypothesis (based on symmetry considerations) 
is that to each LH flavour 
neutrino field $\nu_{lL}(x)$ there corresponds a RH neutrino 
field $\nu_{lR}(x)$, $l=e,\mu,\tau$, 
although schemes with less (more) than three 
RH neutrinos are also being considered (see, e.g., 
\cite{01-Thnu07}).

  The experiments with solar,
atmospheric, reactor and accelerator neutrinos 
(see~\cite{01-PDG12} and the references quoted therein)
have provided compelling evidences for 
flavour neutrino oscillations~\cite{01-BPont67,01-BPont57,01-MNS62} -
transitions in flight between 
the different flavour neutrinos
$\nu_e$, $\nu_\mu$, $\nu_\tau$ 
(antineutrinos $\bar{\nu}_e$, $\bar{\nu}_\mu$, $\bar{\nu}_\tau$),
caused by nonzero neutrino masses and neutrino mixing.
As a consequence of the results of these experiments 
the existence of oscillations 
of the solar $\nu_e$, atmospheric $\nu_{\mu}$ and 
$\bar{\nu}_{\mu}$, 
accelerator $\nu_{\mu}$ (at $L\sim 250$ km, $L\sim 730$ km 
and $L\sim 295$ km,
$L$ being the distance traveled by the neutrinos)
and reactor $\bar{\nu}_e$ (at $L\sim 180$ km and 
$L\sim 1$ km), was firmly established.
The data imply the presence of neutrino mixing 
in the weak charged  lepton current:
\begin{equation}
\label{eq-01:CC}
{\cal L}_{\rm CC} = - ~\frac{g}{\sqrt{2}}\,
\sum_{l=e,\mu,\tau}
\overline{l_L}(x)\, \gamma_{\alpha} \nu_{lL}(x)\,
W^{\alpha \dagger}(x) + h.c.\,,~~
\nu_{l \mathrm{L}}(x)  = \sum^n_{j=1} U_{l j} \nu_{j \mathrm{L}}(x), 
\end{equation}
\noindent where 
$\nu_{lL}(x)$ are the flavour neutrino fields, 
$\nu_{j \mathrm{L}}(x)$ is the left-handed (LH)
component of the field of 
the neutrino $\nu_j$ having a mass $m_j$, 
and $U$ 
is a unitary matrix - the
Pontecorvo-Maki-Nakagawa-Sakata (PMNS)
neutrino mixing matrix~\cite{01-BPont67,01-BPont57,01-MNS62}, 
$U\equiv U_{PMNS}$.
All compelling  neutrino oscillation data
can be described assuming 3-neutrino mixing in vacuum, $n=3$.
The number of massive neutrinos $n$ 
can, in general, be bigger than 3
if, e.g.
there exist right-handed (RH)
sterile neutrinos 
and they mix with the LH flavour neutrinos. 
It follows from the current 
data that at least 3 of 
the neutrinos $\nu_j$, 
say $\nu_1$, $\nu_2$, $\nu_3$, must be light,
$m_{1,2,3} \ltap 1$ eV, and must have different 
masses, $m_1\neq m_2 \neq m_3$. 
At present there are no compelling 
experimental evidence for the existence  
of more than 3 light neutrinos.
Certain neutrino oscillation 
data exhibit anomalies that 
could be interpreted 
as being due to the existence of 
one or two additional (sterile) 
neutrinos with mass in the eV range, 
which have a relatively small mixing 
$\sim 0.1$ with the active flavour 
neutrinos (see, e.g.,~\cite{01-sterile} 
and the references quoted therein.)

 In the case of 3 light neutrinos 
we will concentrate on in this Section, 
the neutrino mixing matrix $U$ can be 
parametrised by 3 angles and, depending on whether 
the  massive neutrinos $\nu_j$ are Dirac 
or Majorana \cite{01-Maj1937} particles, 
by 1 or 3 CP violation (CPV) phases 
~\cite{01-BHP80,01-SchechValle80,01-JBernaPascMajP83}: 
\begin{equation}
U= VP\,,~~~
P = {\rm diag}(1, e^{i \frac{\alpha_{21}}{2}}, e^{i \frac{\alpha_{31}}{2}})\,, 
\label{01-VP}
\end{equation}
%
where $\alpha_{21}$ and $\alpha_{31}$ are the 
two Majorana CPV 
phases and $V$ is a CKM-like matrix 
containing the Dirac CPV 
phase $\delta$,
\begin{equation} 
\begin{array}{c}
\label{01-eq:Vpara}
V = \left(\begin{array}{ccc} 
 c_{12} c_{13} & s_{12} c_{13} & s_{13} e^{-i \delta}  \\[0.2cm] 
 -s_{12} c_{23} - c_{12} s_{23} s_{13} e^{i \delta} 
 & c_{12} c_{23} - s_{12} s_{23} s_{13} e^{i \delta} 
 & s_{23} c_{13} 
\\[0.2cm] 
 s_{12} s_{23} - c_{12} c_{23} s_{13} e^{i \delta} & 
 - c_{12} s_{23} - s_{12} c_{23} s_{13} e^{i \delta} 
 & c_{23} c_{13} 
\\ 
  \end{array} 
\right)\,. 
\end{array} 
\end{equation}
%
\noindent 
In eq.~(\ref{01-eq:Vpara}),
$c_{ij} = \cos\theta_{ij}$, 
$s_{ij} = \sin\theta_{ij}$,
the angles $\theta_{ij} = [0,\pi/2]$, 
$\delta = [0,2\pi]$ 
and, in general, 
$0\leq \alpha_{j1}/2\leq 2\pi$, $j=2,3$~
\cite{01-EMSPEJP09}. 
If CP invariance holds, we have 
$\delta =0,\pi$, and \cite{01-LW81,01-JBernaPascMajP83},
$\alpha_{21(31)} = k^{(')}\,\pi$, $k^{(')}=0,1,2,3,4$.

 Thus, in the case of massive Dirac neutrinos, 
the neutrino mixing matrix $U$ is similar,
in what concerns the number of 
mixing angles and CPV phases, to the CKM quark 
mixing matrix. The presence of two additional 
physical CPV phases in $U$ if $\nu_j$ 
are Majorana particles is a consequence of 
the special properties  of the latter
(see, e.g. refs.~\cite{ 01-BHP80,01-BiPet87}).
On the basis of the existing neutrino data 
it is impossible to determine whether the massive 
neutrinos are Dirac or Majorana fermions.

 The  neutrino oscillation probabilities depend, 
in general, on the neutrino energy, $E$, 
the source-detector distance $L$,
on the elements of 
$U$ and, for relativistic neutrinos used in all neutrino 
experiments performed so far, on the neutrino mass squared 
differences $\Delta m^2_{ij} \equiv (m^2_{i} - m^2_j)$, $i\neq j$ 
(see, e.g., ref.~\cite{01-PDG12,01-BiPet87}).
In the case of 3-neutrino mixing
there are only two independent neutrino 
mass squared differences, 
say $\Delta m^2_{21}\neq 0$ and $\Delta m^2_{31} \neq 0$.
The numbering of massive neutrinos $\nu_j$ is arbitrary.
We will employ here the widely used convention 
of numbering of $\nu_j$ which allows 
to associate $\theta_{13}$ with the smallest 
mixing angle in the PMNS matrix $U$,
and  $\theta_{12}$, $\Delta m^2_{21}> 0$, and
$\theta_{23}$, $\Delta m^2_{31(32)}$,
with the parameters which drive, respectively, 
the solar 
\footnote{Under the assumption of CPT invariance, 
which we will suppose to hold throughout 
this article, $\theta_{12}$ and $\Delta m^2_{21}$
drive also the reactor $\bar{\nu}_e$ 
oscillations at $L\sim 180$ km (see, e.g., \cite{01-PDG12}).}
($\nu_e$), and the dominant atmospheric 
$\nu_{\mu}$ (and $\bar{\nu}_{\mu}$)
(and accelerator $\nu_{\mu}$)  
oscillations. In this convention $m_1 < m_2$,
$ 0 < \Delta m^2_{21} < |\Delta m^2_{31}|$, 
and, depending on ${\rm sgn}(\Delta m^2_{31})$,
we have either $m_3 < m_1$ or $m_3 > m_2$ (see further).
In the case of $m_1 < m_2 < m_3$ ($m_3 < m_1 < m_2$),
the neutrino mass squared 
difference $\Delta m^2_{21}$, 
as it follows from the data to be discussed below, 
is much smaller than $|\Delta m^2_{31(32)}|$,
$\Delta m^2_{21} \ll |\Delta m^2_{31(32)}|$.
This implies that in each of the two cases 
 $m_1 < m_2 < m_3$ and $m_3 < m_1 < m_2$ 
we have $|\Delta m^2_{31}-\Delta m^2_{32}| = 
\Delta m^2_{21} \ll |\Delta m^2_{31,32}|$. 
The angles $\theta_{12}$ and $\theta_{23}$ are sometimes 
called ``solar'' and ``atmospheric'' neutrino mixing 
angles, and are often denoted as 
$\theta_{12} = \theta_{\odot}$ and 
$\theta_{23} = \theta_{\rm atm}$,
while $\Delta m^2_{21}$ and $\Delta m^2_{31(32)}$ 
are sometimes referred to as the ``solar'' and 
``atmospheric'' neutrino mass squared 
differences and are sometimes denoted as~
\footnote{The preceding part of the text of the present 
article follows closely parts of the text of the review 
article \cite{01-PDG12}.} 
$\Delta m^2_{21} \equiv \dmsol$,
$\Delta m^2_{31(32)} \equiv \Delta m^2_{\rm atm}$.

  The neutrino oscillation data, accumulated 
over many years, allowed to determine the parameters 
which drive the solar, reactor,  
atmospheric and accelerator neutrino oscillations, 
$\Delta m^{2}_{21}$, $\theta_{12}$,
$|\Delta m^{2}_{31(32)}|$ 
and  $\theta_{23}$, with a rather high 
precision (see, e.g., \cite{01-Nu2012}).
Furthermore, there were  spectacular developments 
in the period since June 2011 
in what concerns the angle $\theta_{13}$ 
(see, e.g., \cite{01-PDG12}). They culminated 
in March of 2012 in a high precision 
determination of $\sin^22\theta_{13}$ 
in the Daya Bay experiment with reactor 
$\bar{\nu}_e$ 
\cite{01-An:2012eh}:
\begin{equation}
 \sin^22\theta_{13} = 0.089 \pm 0.010 \pm 0.005\,.
\label{01-DBayth13}
\end{equation}
%
Subsequently
\footnote{We have reported in eq. (\ref{01-DBayth13}) 
the latest result of the Daya Bay experiment, published 
in the second article quoted in \cite{01-An:2012eh}.}, 
the RENO \cite{01-RENOth13}, 
Double Chooz, 
and T2K 
experiments \cite{01-DCT2Kth13} 
reported, respectively, $4.9\sigma$, $2.9\sigma$ and 
$3.2\sigma$ evidences for a non-zero value of $\theta_{13}$, 
compatible with the Day Bay result.

  A global analysis of the latest  
neutrino oscillation data presented at 
the Neutrino 2012 International Conference \cite{01-Nu2012}, 
was performed in \cite{01-Fogli:2012XY}. 
We give below the best fit values of $\Delta m^2_{21}$, 
$\sin^2\theta_{12}$, $|\Delta m^2_{31(32)}|$, $\sin^2\theta_{23}$  
and $\sin^2\theta_{13}$, obtained in~\cite{01-Fogli:2012XY}:
\begin{eqnarray}
\label{01-Delta2131}
\Delta m^2_{21} = 7.54 \times 10^{-5}~{\rm eV^2}\,, 
 & |\Delta m^2_{31(32)}| = 2.47~(2.46) \times 10^{-3}~{\rm eV^2}\,,\\ 
 \label{01-sinsq1213}
 \sin^2\theta_{12} = 0.307\,,~~~\sin^2\theta_{23} = 0.39\,,
 & \sin^2\theta_{13} = 0.0241~(0.0244)\,,
 \end{eqnarray}
%
where the values (the values in brackets) correspond to 
$m_1 < m_2 < m_3$ ($m_3 < m_1 < m_2$).
The $1\sigma$ uncertainties and the the $3\sigma$ 
ranges of the neutrino oscillation parameters 
found in~\cite{01-Fogli:2012XY} are given 
\footnote{Note that we have quoted the value 
of $|\Delta m^2_{31(32)}|$ in eq. (\ref{01-Delta2131}),
while the mass squared difference 
determined in~\cite{01-Fogli:2012XY} is 
$|\Delta m^{2}_{A}| = |\Delta m^2_{31} -\Delta m^2_{21}/2|$
($|\Delta m^{2}_{A}| = |\Delta m^2_{32} + \Delta m^2_{21}/2|$).
}
in Table \ref{01-tab:tabdata-0512}.
\begin{table} [t] 
\centering \caption{
\label{01-tab:tabdata-0512} The best-fit values and
$3\sigma$ allowed ranges of the 3-neutrino oscillation parameters,
derived in~\cite{01-Fogli:2012XY} from a global fit of 
the current neutrino oscillation data.
The values (values in brackets) correspond to 
$m_1 < m_2 < m_3$ ($m_3 < m_1 < m_2$). 
The definition of $\Delta m^{2}_{A}$
used is: $\Delta m^{2}_{A} = m^2_3 - (m^2_2 + m^2_1)/2$.
Thus, $\Delta m^{2}_{A} = 
\Delta m^2_{31} -\Delta m^2_{21}/2$, if $m_1 < m_2 < m_3$, 
and  $\Delta m^{2}_{A} = 
\Delta m^2_{32} + \Delta m^2_{21}/2$ for $m_3 < m_1 < m_2$.
}
\renewcommand{\arraystretch}{1.1}
\begin{tabular}{lcc}
\hline \hline
 Parameter  &  best-fit ($\pm 1\sigma$) & 3$\sigma$ \\ \hline
 $\Delta m^{2}_{\odot} \; [10^{-5}\eV^2]$   & 7.54$^{+0.26}_{-0.22}$ &
               6.99 - 8.18 \\
$ |\Delta m^{2}_{A}| \; [10^{-3}\eV^2]$ & 2.43$^{+0.06}_{-0.10}$~
(2.42$^{+0.07}_{-0.11}$)&
           2.19(2.17) - 2.62(2.61)\\
$\sin^2\theta_{12}$  & 0.307$^{+0.018}_{-0.016}$
            & 0.259 - 0.359\\
 $\sin^2\theta_{23}$  & 0.386$^{+0.024}_{-0.021}$
(0.392$^{+0.039}_{-0.022}$)&  0.331(0.335) - 0.637(0.663) \\
$\sin^2\theta_{13}$ &  0.0241$\pm 0.0025$ 
(0.0244$^{+0.0023}_{-0.0025}$)& 0.0169(0.0171) - 0.0313(0.0315) \\
\hline\hline
\end{tabular}
\end{table}
%

 A few comments are in order.
We have $\Delta m^2_{21}/|\Delta m^2_{31(32)}| \cong 0.031 \ll 1$, 
as was indicated earlier. 
The existing data do not allow to determine the 
sign of $\Delta m^2_{31(32)}$. As we will discuss 
further, the two possible signs correspond to 
two different basic types of neutrino 
mass spectrum. Maximal solar neutrino mixing, i.e. 
$\theta_{12} = \pi/4$, is ruled out at more than 
6$\sigma$ by the data. Correspondingly, one has  
$\cos2\theta_{12} \geq 0.28$ at $3\sigma$.
The results quoted in eq. (\ref{01-sinsq1213}) 
imply that $\theta_{23} \cong \pi/4$, $\theta_{12} \cong \pi/5.4$ and 
that $\theta_{13} \cong \pi/20$.
Thus, the pattern of 
neutrino mixing is drastically different 
from the pattern of quark mixing. 
As we have noticed earlier,
the neutrino oscillations 
experiments are sensitive only to 
neutrino mass squared differences 
$\Delta m^2_{ij} \equiv (m^2_{i} - m^2_j)$, $i\neq j$,   
and cannot give information on the
absolute values of the neutrino masses, 
i.e., on the absolute neutrino mass scale.
They are insensitive also
to the nature - Dirac or Majorana, 
of massive neutrinos $\nu_j$ and, 
correspondingly, to the Majorana CPV 
phases present in the PMNS matrix $U$
~\cite{01-BHP80,01-Lang87}.

 After the successful measurement of $\theta_{13}$,
the determination of  the absolute neutrino mass scale, 
of the type of the neutrino mass spectrum, 
of the nature - Dirac or Majorana, 
of massive neutrinos, 
as well as getting information about 
the status of CP violation in the lepton sector, 
are the most pressing and challenging problems 
and the highest priority 
goals of the research in the field of 
neutrino physics (see, e.g., \cite{01-PDG12}).

   As was already indicated above,
the presently available data 
do not permit to determine the sign of 
$\Delta m^2_{31(2)}$.
In the case of 3-neutrino mixing,  
the two possible signs of
$\Delta m^2_{31(32)}$ correspond to two 
types of neutrino mass spectrum.
In the widely used convention of numbering 
the neutrinos with definite mass 
employed by us, the two spectra read:\\
{\it i) spectrum with normal ordering (NO)}:
$m_1 < m_2 < m_3$, $\dma = \Delta m^2_{31} >0$,
$\dmsol \equiv \Delta m^2_{21} > 0$,
$m_{2(3)} = (m_1^2 + \Delta m^2_{21(31)})^{1\over{2}}$; \\~~
{\it ii) spectrum with inverted ordering (IO)}:
$m_3 < m_1 < m_2$, $\dma = \Delta m^2_{32}< 0$, 
$\dmsol \equiv \Delta m^2_{21} > 0$,
$m_{2} = (m_3^2 + \Delta m^2_{23})^{1\over{2}}$, 
$m_{1} = (m_3^2 + \Delta m^2_{23} - \Delta m^2_{21})^{1\over{2}}$.\\

\vspace{-0.4cm}
\noindent 
Depending on the value of the lightest neutrino mass, 
${\rm min}(m_j)$, the neutrino mass spectrum can be:\\
%
{\it a) Normal Hierarchical (NH)}: 
$m_1 \ll m_2 < m_3$, $m_2 \cong (\dmsol)^{1\over{2}} 
\cong 8.68 \times 10^{-3}$ eV,
$m_3 \cong |\dma|^{1\over{2}} \cong 4.97\times 10^{-2}~{\rm eV}$; or  \\
%
{\it b) Inverted Hierarchical (IH)}: $m_3 \ll m_1 < m_2$, 
with $m_{1,2} \cong |\dma|^{1\over{2}}\cong 4.97\times 10^{-2}$ eV; or  \\ 
%
{\it c) Quasi-Degenerate (QD)}: $m_1 \cong m_2 \cong m_3 \cong m_0$,
$m_j^2 \gg |\dma|$, $m_0 \gtap 0.10$ eV. \\ 
%

  The type of neutrino mass spectrum (hierarchy), i.e., 
the sign of $\Delta m^2_{31(32)}$,
can be determined i) using data from
neutrino oscillation experiments 
at accelerators (NO$\nu$A, T2K, etc.) 
(see, e.g., \cite{01-Future}), 
ii) in the experiments studying the oscillations 
of atmospheric neutrinos (see, e.g., \cite{01-2JBSP203}), 
as well as iii) in experiments with reactor antineutrinos
\cite{01-ReactNuHiera}. The relatively large value of 
$\theta_{13}$ is a favorable factor for
the ${\rm sgn}(\Delta m^2_{31(32)})$ determination
in these experiments.
If neutrinos with definite mass are
Majorana particles, information about 
the ${\rm sgn}(\Delta m^2_{31(32)})$ can be obtained 
also by measuring the effective neutrino Majorana
mass in neutrinoless double $\beta-$decay 
experiments \cite{01-PPSNO2bb}.

 More specifically, in the cases i) and ii)
the ${\rm sgn}(\Delta m^2_{31(32)})$ can be determined 
by studying oscillations of neutrinos and
antineutrinos, say, $\nu_{\mu} \leftrightarrow \nu_e$
and $\bar{\nu}_{\mu} \leftrightarrow \bar{\nu}_e$,
in which matter effects are sufficiently large.
This can be done in long base-line 
$\nu$-oscillation experiments 
(see, e.g., \cite{01-Future}). 
For $\sin^22\theta_{13}\gtap 0.05$
and $\sin^2\theta_{23}\gtap 0.50$,
information on ${\rm sgn}(\Delta m^2_{31(32)})$
might be obtained in atmospheric 
neutrino experiments by investigating 
the effects of the sub-dominant transitions
$\nu_{\mu(e)} \rightarrow \nu_{e(\mu)}$
and $\bar{\nu}_{\mu(e)} \rightarrow \bar{\nu}_{e(\mu)}$ 
of atmospheric  neutrinos which traverse 
the Earth (for a detailed discussion see, e.g., \cite{01-2JBSP203}). 
For $\nu_{\mu(e)}$ ({\it or} $\bar{\nu}_{\mu(e)}$) 
crossing the Earth core, new type of resonance-like
enhancement of the indicated transitions
takes place due to the {\it (Earth) mantle-core
constructive interference effect
(neutrino oscillation length resonance (NOLR))} 
\cite{01-SP3198} (see also \cite{01-ChMaris98}). 
As a consequence of this effect
the corresponding 
$\nu_{\mu(e)}$ ({\it or} $\bar{\nu}_{\mu(e)}$)
transition probabilities can be maximal 
\cite{01-106107} (for the precise conditions
of the mantle-core (NOLR) enhancement
see \cite{01-SP3198,01-106107})
\footnote{
We note that the Earth mantle-core (NOLR) 
enhancement of neutrino transitions 
differs \cite{01-SP3198} from the MSW one. 
It also differs \cite{01-SP3198,01-106107} from the 
mechanisms of enhancement 
discussed, e.g., in the articles \cite{01-Param86}: 
the conditions of 
enhancement considered in  \cite{01-Param86}
cannot be realised for the 
$\nu_{\mu(e)} \rightarrow \nu_{e(\mu)}$
or $\bar{\nu}_{\mu(e)} \rightarrow \bar{\nu}_{e(\mu)}$ 
transitions of the Earth core crossing
neutrinos.
}. 
For $\Delta m^2_{31(32)}> 0$, the neutrino transitions
$\nu_{\mu(e)} \rightarrow \nu_{e(\mu)}$
are enhanced, while for $\Delta m^2_{31(32)}< 0$
the enhancement of antineutrino transitions
$\bar{\nu}_{\mu(e)} \rightarrow \bar{\nu}_{e(\mu)}$
takes place \cite{01-SP3198} 
(see also \cite{01-ChMaris98,01-106107,01-AkhMaltSmir06}), 
which might allow to determine 
\footnote{The so-called ``neutrino oscillograms of the Earth'', 
showing in the case of $\Delta m^2_{31(32)}> 0$, 
for instance, the dependence of, e.g., the  
$\nu_{\mu(e)} \rightarrow \nu_{e(\mu)}$ 
transition probability 
on the nadir angle of the neutrino trajectory 
through the Earth for given values of 
the neutrino energy $E$ (or  $\Delta m^2_{31(32)}/E$) 
and of the relevant neutrino mixing angle,
and discussed in detail in \cite{01-AkhMaltSmir06} 
and in the talks given by the authors of 
\cite{01-AkhMaltSmir06}, first appeared in the 
publications quoted in \cite{01-ChMaris98}.}
${\rm sgn}(\Delta m^2_{31(32)})$. 
Determining the type of neutrino mass spectrum 
is crucial for understanding the origin of 
neutrino masses and mixing as well.

  All possible types of neutrino mass spectrum 
we have discussed above are compatible
with the existing constraints on the absolute scale 
of neutrino masses $m_j$. 
Information about the absolute neutrino mass scale
can be obtained by measuring the spectrum
of electrons near the end point in $^3$H $\beta$-decay 
experiments \cite{01-Fermi34}
and from cosmological and astrophysical data 
(see, e.g., \cite{01-Cosmo}).
The most stringent upper
bound on the $\bar{\nu}_e$ mass
was obtained in the Troitzk~\cite{01-MoscowH3b}
experiment (see also~\cite{01-MainzKATRIN}):
\beq
m_{\bar{\nu}_e} < 2.05~\rm{eV}~~~\mbox{at}~95\%~\rm{C.L.} 
\label{01-H3beta}
\eeq
%
\noindent We have $m_{\bar{\nu}_e} \cong m_{1,2,3}\gtap 0.1$ eV
in the case of quasi-degenerate (QD)
spectrum. The KATRIN experiment~\cite{01-MainzKATRIN}, which 
is under preparation, is planned to reach 
sensitivity of  $m_{\bar{\nu}_e} \sim 0.20$~eV,
i.e., it will probe the region of the QD spectrum
\footnote{Information on the type of neutrino mass spectrum
can also be obtained in $\beta$-decay experiments 
having a sensitivity to neutrino masses  
$\sim \sqrt{|\Delta m^2_{31(32)}|}\cong 5\times 10^{-2}$ eV 
\cite{01-BMP06} (i.e., by a factor of $\sim 4$ 
better sensitivity than that of the KATRIN 
experiment \cite{01-MainzKATRIN}). Reaching the 
indicated sensitivity in electromagnetic spectrometer 
$\beta$-decay experiments of the type of 
KATRIN does not seem feasible at present.
}. 
The Cosmic Microwave Background (CMB)
data of the WMAP experiment, combined with
supernovae data and data on galaxy clustering
can be used to derive an upper limit on the sum of
neutrinos masses (see, e.g., \cite{01-Cosmo}). 
Depending on the model complexity and the input data used 
one obtains~\cite{01-summj}:
$\sum_j m_j\ltap (0.3 - 1.3)$ eV, 95\% C.L.
Data on weak lensing of galaxies, 
combined with data from the WMAP and PLANCK
experiments, may allow $\sum_j m_j$ 
to be determined with an uncertainty of 
$\sigma(\sum_j m_j) =
(0.04 - 0.07)$~eV~\cite{01-Hu2005}.

 Thus, the data on the 
absolute scale of neutrino masses  
imply that neutrino masses are much smaller than 
the masses of the charged leptons and quarks.
If we take as an indicative upper limit $m_j \ltap 0.5$ eV, 
$j=1,2,3$, we have
\beq
\frac{m_j}{m_{l,q}} \ltap 10^{-6}\,,~~l\, =\, e,\mu,\tau\,,~~
{\rm q\,= \,d,s,b,u,c,t}\,. 
\label{01-mjovermlq}
\eeq
%
It is natural to suppose that the remarkable
smallness of neutrino masses is related to 
the existence of a new fundamental mass scale
in particle physics, and thus to new physics beyond 
that predicted by the Standard Model.
A comprehensive theory of the neutrino 
masses and mixing should be able to explain 
the indicated enormous  disparity between the neutrino 
masses and the masses of the charged 
leptons and quarks.

  At present no experimental information on 
the Dirac and Majorana CPV 
phases in the neutrino mixing matrix is available.
Therefore the status of the 
CP symmetry in the lepton sector is unknown. 
 The importance of getting information 
about the Dirac and Majorana CPV phases 
in the neutrino mixing matrix 
stems, in particular, from the
possibility that these phases 
play a fundamental 
role in the generation of the 
observed baryon asymmetry of the Universe.
More specifically, the CP violation 
necessary for the generation of the 
baryon asymmetry within the ``flavoured'' 
leptogenesis scenario \cite{01-FLG1,01-FLG2}
can be due exclusively 
to the Dirac and/or Majorana CPV phases 
in the PMNS matrix \cite{01-PPRio106},
and thus can be 
directly related to the 
low energy CP-violation in 
the lepton sector. 
If the requisite CP violation is due to
the Dirac phases $\delta$,
a necessary condition for a 
successful (''flavoured'') leptogenesis
is that $\sin\theta_{13} \gtap 0.09$ 
\cite{01-PPRio106}, which is comfortably
compatible with the Daya Bay result, eq. (\ref{01-DBayth13}).

 With $\theta_{13} \neq 0$,
the Dirac 
phase $\delta$ can generate
CP violating effects in neutrino oscillations 
\cite{01-Cabibbo78} 
(see also \cite{01-BHP80,01-VBarg80CP}),
i.e., a difference between the probabilities of 
$\nu_l \rightarrow \nu_{l'}$ and 
$\bar{\nu}_l \rightarrow \bar{\nu}_{l'}$
oscillations in vacuum: 
$P(\nu_l \rightarrow \nu_{l'}) \neq 
P(\bar{\nu}_l \rightarrow \bar{\nu}_{l'})$, 
$l\neq l' = e,\mu,\tau$.
The magnitude of the  CP violating 
effects of interest
is determined \cite{01-PKSP3nu88}
by the rephasing invariant 
$J_{CP}$ associated with the Dirac 
CPV 
phase $\delta$ in $U$.
It is analogous to the rephasing invariant 
associated with the Dirac CPV 
phase in the CKM quark mixing 
matrix \cite{01-CJ85}. In the ``standard'' 
parametrisation 
of the PMNS neutrino mixing 
matrix, eqs.~ (\ref{01-VP}) - (\ref{01-eq:Vpara}), 
we have:
\beq
J_{CP} \equiv 
{\rm Im}\,(U_{\mu 3}\,U^*_{e3}\,U_{e2}\,U^*_{\mu 2}) = 
\frac{1}{8}\cos\theta_{13}
\sin 2\theta_{12}\sin 2\theta_{23}\sin 2\theta_{13}\sin \delta\,.
\label{01-JCPstpar}
\eeq
%
Thus, given the fact that 
$\sin \theta_{12}$, $\sin \theta_{23}$ and 
$\sin \theta_{13}$ have been determined experimentally with 
a relatively good precision, 
the size of CP violation effects in neutrino oscillations 
depends essentially only on the magnitude of the currently 
unknown value of the Dirac phase $\delta$. 
The current data imply 
$|J_{CP} | \ltap 0.039$,
where we have used eq. (\ref{01-JCPstpar}) and the $3\sigma$
ranges of $\sin^2\theta_{12}$, $\sin^2\theta_{23}$ and 
$\sin^2\theta_{13}$ given in Table \ref{01-tab:tabdata-0512}.
Data on the Dirac phase $\delta$
will be obtained in the long baseline 
neutrino oscillation experiments 
T2K, NO$\nu$A, etc.
(see, e.g., refs.~\cite{01-PDG12}).
Testing the possibility of Dirac 
CP violation in the lepton sector is one of 
the major goals of the next generation of 
neutrino oscillation experiments 
(see, e.g., \cite{01-Future,01-JBerna2010}).
Measuring the magnitude of CP violation effects 
in neutrino oscillations 
is at present also the only known feasible
method of determining the value of the 
phase $\delta$ (see, e.g., \cite{01-Branco2012}).  

  If $\nu_j$ are Majorana fermions,
getting experimental information about the  
Majorana CPV phases in the neutrino mixing 
matrix $U$ will be remarkably difficult
~\cite{01-BPP1,01-PPW,01-PPR1,01-PPSchw05,01-BargerCP,01-MajPhase1,
01-LisiFess08}. 
As we will discuss further, the Majorana phases 
of the PMNS matrix play important role 
in the phenomenology of neutrinoless double beta 
($\betabeta$-) decay - the process whose 
existence is related to the Majorana nature 
of massive neutrinos \cite{01-Racah1937}:
$(A,Z) \rightarrow (A,Z+2) + e^- + e^-$. 
The phases $\alpha_{21,31}$ 
can affect significantly the predictions for 
the rates of the (LFV) decays $\mu \rightarrow e + \gamma$,
$\tau \rightarrow \mu + \gamma$, etc.
in a large class of supersymmetric theories
incorporating the see-saw mechanism 
\cite{01-PPY03}. 
As was mentioned earlier,
the Majorana phase(s) in the PMNS matrix can
be the leptogenesis CP violating parameter(s) 
at the origin of the baryon asymmetry of the Universe. 
\cite{01-PPRio106,01-EMSPEJP09}.

 Establishing whether the neutrinos with definite mass
$\nu_j$ are Dirac fermions possessing distinct antiparticles, 
or Majorana fermions, i.e., spin 1/2 particles that 
are identical with their antiparticles, is 
of fundamental importance for understanding 
the origin of $\nu$-masses and mixing and 
the underlying symmetries of particle 
interactions.
Let us recall that the neutrinos $\nu_j$
with definite mass $m_j$ will be Dirac fermions if 
particle interactions conserve 
some additive lepton number, e.g., the total
lepton charge $L = L_e + L_{\mu} + L_{\tau}$. 
If no lepton charge is conserved, 
the neutrinos $\nu_j$ will be Majorana fermions 
(see, e.g. ~\cite{01-BiPet87}). 
The massive neutrinos are 
predicted to be of Majorana nature
by the see-saw mechanism 
of neutrino mass generation 
\cite{01-seesaw}, which also provides an  
attractive explanation of the
smallness of neutrino masses 
and, through the leptogenesis theory 
\cite{01-LG}, of the observed baryon 
asymmetry  of the Universe.
The observed patterns of 
neutrino mixing 
and of neutrino mass squared differences
driving the solar and the dominant 
atmospheric neutrino oscillations,
can be related to Majorana massive neutrinos 
and the existence of an 
{\it approximate} symmetry in the lepton sector
corresponding to the conservation of the
{\it non-standard} lepton charge
$L' = L_e - L_{\mu} - L_{\tau}$ \cite{01-STPPD82}.
They can also be associated with the existence 
of {\it approximate} discrete symmetry 
(or symmetries) of the particle interactions 
(see, e.g., \cite{01-Descrete}).
Determining the nature (Dirac or Majorana)
of massive neutrinos is one of the fundamental 
and most challenging  
problems in the future studies of neutrino mixing
\cite{01-PDG12}.

\section{The Nature of Massive Neutrinos}

\subsection{Majorana versus Dirac Massive Neutrinos (Particles)} 
\label{sec-01:second} 

 The properties of Majorana particles (fields) are
very different from those of Dirac particles (fields).
A massive Majorana neutrino $\chi_j$ 
(or Majorana spin 1/2 particle)
with mass $m_j > 0$ 
can be described in local quantum field theory 
which is used to construct, e.g., the Standard Model,
by 4-component complex spin 1/2 field $\chi_j(x)$
which satisfies the Dirac equation and 
the Majorana condition:
\beq
C~(\bar{\chi}_j(x))^{{\rm T}} = \xi_k \chi_j(x)\,, ~~|\xi_j|^2 = 1\,,
\label{01-MajCon}
\eeq
%
\noindent where $C$ is the charge conjugation matrix, 
$C^{-1}\,\gamma_{\alpha}\,C = -\,(\gamma_{\alpha})^{T}$ 
($C^T = -\,C$, $C^{-1} = C^{\dagger}$), 
and  $\xi_j$ is, in general, an unphysical phase.
The Majorana condition is invariant under {\it proper} 
Lorentz transformations. It reduces by 
a factor of 2 the number of independent components in $\chi_j(x)$.

  The condition (\ref{01-MajCon}) is invariant with
respect to $U(1)$ global gauge transformations
of the field $\chi_j(x)$ carrying a $U(1)$ charge $Q$,
$\chi_j(x) \rightarrow e^{i\alpha Q}\chi_j(x)$, only if $Q = 0$.
As a result, i) $\chi_j$ cannot carry nonzero 
additive quantum numbers (lepton charge, etc.),
and ii) the field $\chi_j(x)$ cannot ``absorb'' phases.
Thus, $\chi_j(x)$ describes 2 spin states of a spin 1/2, 
{\it absolutely neutral particle}, which is identical with
its antiparticle, $\chi_j \equiv \bar{\chi}_j$.
As is well known, spin 1/2 Dirac particles can carry 
nonzero $U(1)$ charges: the charged leptons and quarks, 
for instance, carry nonzero electric charges.

  Owing to the fact that the Majorana (neutrino) fields 
cannot absorb phases, the neutrino mixing matrix $U$ 
contains in the general case of $n$ charged leptons 
and mixing of $n$ massive Majorana 
neutrinos $\nu_j\equiv \chi_j$, altogether 
\beq
n^{\rm (M)}_{\rm CPV} = \frac{n(n-1)}{2}\,,~~~{\rm Majorana}~\nu_j\,,
\label{01-nCPVM} 
\eeq
%
CPV phases~\cite{01-BHP80}. In the case of mixing of $n$ massive Dirac 
neutrinos, the number of CPV phases in $U$, as is well known, is  
\beq
n^{\rm (D)}_{\rm CPV} = \frac{(n-1)(n-2)}{2}\,,~~~{\rm Dirac}~\nu_j\,.
\label{01-nCPVD} 
\eeq
%
Thus, if $\nu_j$ are Majorana particles, $U$ contains 
the following number of additional {\it Majorana} 
CP violation phases: $n^{\rm (M)}_{\rm MCPV}\equiv  
 n^{\rm (M)}_{\rm CPV} - n^{\rm (D)}_{\rm CPV} = (n-1)$.
In the case of $n$ charged leptons and $n$ massive Majorana neutrinos,
the PMNS matrix $U$ can be cast in the form \cite{01-BHP80}
\beq
U = V \,P\,,
\label{01-VP1}
\eeq
%
\noindent where the matrix $V$ contains
the $(n-1)(n-2)/2$ Dirac CP violation
phases, while $P$ is a diagonal matrix
with the additional  $(n-1)$
Majorana CP violation phases
$\alpha_{21}$, $\alpha_{31}$,..., $\alpha_{n1}$,
\beq
P = diag \left (1,e^{i{{\alpha_{21}}\over{2}}},e^{i{{\alpha_{31}}\over{2}}},...,
e^{i{{\alpha_{n1}}\over{2}}}\right )\,.
\label{01-MajP2}
\eeq
%
As will discuss further, the Majorana phases will conserve CP 
if \cite{01-LW81} $\alpha_{j1} = \pi q_j$, $q_j=0,1,2$, $j=2,3,...,n$.
In this case  ${\rm exp}(i\alpha_{j1}) = \pm 1$ and
 ${\rm exp}[i(\alpha_{j1} - \alpha_{k1})] = \pm 1$
have a simple physical interpretation: these are the relative
CP-parities of the Majorana neutrinos $\nu_j$ and $\nu_1$ and
of $\nu_j$ and $\nu_k$, respectively.

 It follows from the preceding discussion that the mixing 
of massive Majorana neutrinos differs, in what concerns the 
number of CPV phases, from the mixing of 
massive Dirac neutrinos. 
For $n=3$ of interest, we have one Dirac and 
two Majorana CPV phases in $U$, which is consistent 
with the expression of $U$ given in eq. (\ref{01-VP}). 
If $n=2$, there is one Majorana CPV phase and 
no Dirac CPV phases in $U$. 
Correspondingly, in contrast to the Dirac case, 
there can exist CP violating effects 
even in the system of two mixed massive Majorana 
neutrinos (particles).
  
  The Majorana phases do not enter into the expressions of the 
probabilities of oscillations involving the flavour neutrinos 
and antineutrinos \cite{01-BHP80,01-Lang87}, $\nu_l \rightarrow \nu_{l'}$ and 
$\overline{\nu}_l \rightarrow \overline{\nu}_{l'}$. Indeed, 
the probability to find neutrino $\nu_{l'}$ 
(antineutrino  $\overline{\nu}_{l'}$) at time $t$ if a neutrino 
 $\nu_l$ (antineutrino  $\overline{\nu}_{l}$) has been been produced 
at time $t_0$ and it had traveled a distance $L\cong t$ in vacuum,  
is given by (see, e.g., \cite{01-BiPet87,01-PDG12}):
\begin{eqnarray}
\label{01-Pnu}
P(\nu_l\rightarrow \nu_{l'})
&=& \left | \sum_{j} U_{l'j}\,e^{-i(E_j\,t - p_j\,L)}\,
U^{\dagger}_{jl}\right|^2\,,\\
P(\overline{\nu}_l \rightarrow \overline{\nu}_{l'}) = 
&=& \left |\sum_{j} U_{lj}\,e^{-i(E_j\,t - p_j\,L)}\,
U^{\dagger}_{jl'}\right|^2\,,
\label{01-Pantinu}
\end{eqnarray}
%
where $E_j$ and $p_j$ are the energy and momentum of the neutrino 
$\nu_j$. It is easy to show, 
using the expression for $U$ in eq. (\ref{01-VP1}) that 
$P(\nu_l\rightarrow \nu_{l'})$ and 
$P(\overline{\nu}_l \rightarrow \overline{\nu}_{l'})$ 
do not depend on the Majorana phases present in $U$ since
\beq
\sum_{j} (VP)_{l'j}\,e^{-i(E_j\,t - p_j\,L)}\,(VP)^{\dagger}_{jl}
= \sum_{j} V_{l'j}\,e^{-i(E_j\,t - p_j\,L)}\,V^{\dagger}_{jl}\,.
\eeq
%
The same result holds when the neutrino oscillations 
take place in matter \cite{01-Lang87}.

  If $CP$-invariance holds, Majorana neutrinos (particles) 
have definite $CP$-parity $\eta_{CP}(\chi_j) = \pm i$:
\beq
U_{CP}~\chi_j(x)~U^{-1}_{CP} = \eta_{CP}(\chi_j)~\gamma_{0}~\chi_j(x_p),~~
\eta_{CP}(\chi_j) \equiv i\rho_j = \pm i~\,,
\label{01-CPMaj} 
\eeq
%
where $x = (x_0,{\bf x})$, $x_p = (x_0,-{\bf x})$ 
and $U_{\rm CP}$ is the unitary CP-transformation operator. 
In contrast, Dirac particles do not have a definite 
$CP$-parity $-$ a Dirac field $f(x)$ transforms as follows 
under the CP-symmetry operation:
\beq
U_{CP}\,f(x)\,U^{-1}_{CP} =  
\eta_{f}\, \gamma_0\, C(\overline{f}(x_p))^{T}\,,~~|\eta_{f}|^2=1 \,,
\label{01-CPDirac} 
\eeq
%
$\eta_{f}$ being an unphysical phase factor.
In the case of $CP$ invariance, 
the $CP$-parities of massive Majorana 
fermions (neutrinos) can play important role in 
processes involving real of virtual Majorana particles 
(see, e.g., \cite{01-BiPet87,01-eechi1chi286,01-SUSYMajCP}).

  Using  eqs. (\ref{01-CPDirac}) and  (\ref{01-CPMaj}) and 
the transformation of the $W^{\pm}$ boson 
field under the CP-symmetry operation,
\beq
U_{\rm CP}\, W_{\alpha}(x)\, U^{\dagger}_{\rm CP}
= \eta_{W}\,\kappa_{\alpha}(W_{\alpha}(x_p))^{\dagger}\,,~|\eta_{W}|^2=1\,,~
\kappa_{0} = -1\,,\kappa_{1,2,3}=+1\,.
\label{01-CPW}
\eeq
%
where $\eta_{W}$ is an unphysical phase, 
one can derive the constraints on the 
neutrino mixing matrix $U$ following 
from the requirement of CP-invariance 
of the leptonic CC weak interaction Lagrangian, 
eq. (\ref{eq-01:CC}).
In the case of massive Dirac neutrinos we obtain:
$\eta^*_{\nu_j}\eta_{l} \eta_{W}U^*_{lj} = U_{lj}$, 
$l=e,\mu,\tau$, $j=1,2,3$. 
Setting the product of unphysical phases
$\eta^*_{\nu_j}\eta_{l} \eta_{W} = 1$,
one obtains the well known result: 
\beq
{\rm CP~invariance}:~~U^*_{lj} = U_{lj}\,,~l=e,\mu,\tau\,,~j=1,2,3\,,
~~~{\rm Dirac}~\nu_j\,.
\label{01-CPUDirac}
\eeq
%
In the case of massive Majorana neutrinos we obtain 
using eqs. (\ref{01-MajCon}), (\ref{01-CPMaj})  
(\ref{01-CPDirac}) and (\ref{01-CPW}):
$\xi^*_j (i\rho_{j})\eta^*_{l} \eta_{W} U_{lj} = U^*_{lj}$.
It is convenient now to set $\xi_j =1$,
 $\eta_{l} = i$ and $\eta_{W} = 1$. 
In this (commonly  used by us) convention we get \cite{01-BiPet87}:
\beq
{\rm CP~invariance}:~~U^*_{lj} = \rho_j\,U_{lj}\,,~\rho_j = +1~{\rm or}~(-1)\,,
~l=e,\mu,\tau\,,~j=1,2,3\,,
~~{\rm Majorana}~\nu_j\,.
\label{01-CPUMaj}
\eeq
%
Thus, in the convention used 
the elements of the PMNS matrix can be either 
real or purely imaginary if $\nu_j$ are Majorana fermions.
Applying the above conditions to, e.g., $U_{e2}$, $U_{\tau 3}$ 
and $U_{e 3}$ elements of the PMNS matrix (\ref{01-VP})
we obtain the CP conserving values of the 
phases $\alpha_{21}$, $\alpha_{31}$ and  $\delta$, respectively: 
$\alpha_{21} = k\pi$, $k=0,1,2,...$, 
$\alpha_{31} = k'\pi$, $k'=0,1,2,...$, 
$\delta = 0,\pi,2\pi$.

 One can obtain in a similar way the CP-invariance 
constraint on the matrix of neutrino Yukawa couplings, 
$\lambda_{k l}$, which plays a fundamental 
role in the leptogenesis scenario of baryon asymmetry 
generation, based on the (type I) see-saw mechanism 
of generation of  neutrino masses \cite{01-LG,01-LG1,01-Branco2012}:
\begin{eqnarray}
\label{01-Ynu}
{\cal L}_{\rm Y}(x) &=& 
-\,\lambda_{k l}\,\overline{N_{k R}}(x)\, H^{\dagger}(x)\,\psi_{l L}(x) 
+ \hbox{h.c.}\,,\\ 
\label{01-MN}
{\cal L}_{\rm M}^{\rm N}(x) 
&=& -\,\frac{1}{2}\,M_{k}\,\overline{N_k}(x)\, N_k(x)\,.
\end{eqnarray}
%
Here $N_{kR}(x)$ is the field of a 
heavy right-handed (RH) sterile 
Majorana neutrino with mass $M_k > 0$,
$\psi_{lL}$ denotes 
the Standard Model left-handed (LH) 
lepton doublet field of flavour 
$l=e,\mu,\tau$, $\psi^{\rm T}_{lL} = (\nu^{\rm T}_{lL}~l^{\rm T}_L)$,   
and $H$ is the Standard Model Higgs doublet field whose neutral 
component has a vacuum expectation value 
$v=174$ GeV. The term  ${\cal L}_{\rm Y}(x) 
+ {\cal L}_{\rm M}^{\rm N}(x)$ 
includes all the necessary ingredients of
the see-saw mechanism. Assuming the existence of 
two heavy Majorana neutrinos, i.e., taking $k=1,2$ in 
eqs. (\ref{01-Ynu}) and (\ref{01-MN}), 
and adding the term  ${\cal L}_{\rm Y}(x) 
+ {\cal L}_{\rm M}^{\rm N}(x)$ 
to the Standard Model Lagrangian, we obtain 
the minimal extension of the Standard Model 
in which the neutrinos have masses and mix and
the leptogenesis can be realised. 
In the leptogenesis formalism it is often 
convenient to use the so called ``orthogonal 
parametrisation`` of the matrix 
of neutrino Yukawa couplings \cite{01-Casas:2001sr}:
\begin{equation}
\label{01-R}
\lambda_{k l} = \frac{1}{v} \,  \sqrt{M_k} \, R_{kj}\, 
\sqrt{m_j}\, (U^{\dagger})_{j l}\;,
\end{equation} 
%
\noindent where $R$ is, in general, 
a complex orthogonal matrix, 
$R~R^T = R^T~R = {\bf 1}$.
The CP violation necessary for the generation 
of the baryon asymmetry of the Universe is provided 
in the leptogenesis scenario of interest 
by the matrix of neutrino Yukawa couplings $\lambda$ 
(see, e.g., \cite{01-LG1,01-Branco2012}).
It follows from eq. (\ref{01-R}) that it can be provided
 either by the neutrino mixing matrix $U$, or by the 
matrix $R$, or else by both the matrices 
$U$ and $R$. It is therefore important to derive 
the conditions under which $\lambda$, $R$ and 
$U$ respect the CP symmetry. For the PMNS matrix $U$
these conditions are given in eq. (\ref{01-CPUMaj}).
For the matrices  $\lambda$ and $R$  in the convention 
in which i) $N_{k}(x)$ 
satisfy the Majorana condition with a 
phase equal to 1 (i.e., 
$\xi_k =1$), ii) $\eta^{l} = i$ and 
$\eta^{H} = 1$,  $\eta^{l}$ and $\eta^{H}$
being the unphysical phase factors which appear in the
CP-transformations of the LH lepton doublet  
and Higgs doublet fields 
\footnote{This convention is similar to, and consistent with, 
the convention about the unphysical phases we have used to derive 
the CP-invariance constraints on the elements of 
the PMNS matrix $U$.}, 
$\psi_{lL}(x)$ and $H(x)$, respectively, 
they read \cite{01-PPRio106}:
\begin{equation}
\lambda^{\ast}_{k l} = \lambda_{k l}\,\rho^{N}_k\,,~\rho^{N}_k = \pm 1\,,
~~j=1,2,3,~l=e,\mu,\tau,
\label{01-YCPinv1}
\end{equation}
\begin{equation}
R^{\ast}_{kj} = R_{kj}\,\rho^{N}_k \,\rho_j,~~j,k=1,2,3\,. 
\label{01-RCPinv1}
\end{equation}
%
where $i\rho^{N}_k \equiv \eta_{CP}(N_k) = \pm i$ is the CP-parity 
of $N_k$. Thus, in the case of CP invariance also 
the elements of $\lambda$ and $R$ can be real or purely imaginary.
Note that, as it follows from eqs. (\ref{01-CPUMaj}), 
(\ref{01-YCPinv1}) and (\ref{01-RCPinv1}), 
given which elements are real and which are purely imaginary 
of any two of the three matrices $U$, $\lambda$ and $R$, 
determines in the convention we are using
and if CP invariance holds, 
which elements are real or purely imaginary in the 
third matrix. If, for instance, $U_{e2}$ is purely 
imaginary ($\rho_2 = -1$) and $\lambda_{1 \mu}$ is real ($\rho^{N}_1 = 1$),
then $R_{12}$ must be purely imaginary. Thus, in the example 
we are considering, a real $R_{12}$ would signal that 
the CP symmetry is broken \cite{01-PPRio106}.

 The currents formed by Majorana fields have special properties, 
which make them also quite different from the currents 
formed by Dirac fields. In particular, 
it follows from the Majorana condition that 
the following currents of the Majorana field 
$\chi_{k}(x)$ are identically equal to zero 
(see, e.g., \cite{01-BiPet87}):
\begin{eqnarray}
\label{01-VMaj0}
&&\bar{\chi}_{k}(x)\,\gamma_{\alpha}\,\chi_{k}(x) \equiv 0\,,\\
\label{01-TMaj0}
&&\bar{\chi}_{k}(x)\,\sigma_{\alpha\beta}\,\chi_{k}(x) \equiv 0\,,\\
\label{01-PTMaj0}
&&\bar{\chi}_{k}(x)\,\sigma_{\alpha\beta}\,\gamma_5\,\chi_{k}(x) \equiv 0\,.
\end{eqnarray}
%
Equations (\ref{01-VMaj0}), (\ref{01-TMaj0}) and (\ref{01-PTMaj0})
imply that Majorana fermions (neutrinos)
cannot have nonzero $U(1)$ charges and 
intrinsic magnetic and electric dipole 
moments, respectively.
A Dirac spin 1/2 particle can have non-trivial  
$U(1)$ charges, as we have already discussed, 
and nonzero intrinsic magnetic moment 
(the electron and the muon, for instance, have it).
If CP invariance does not hold, Dirac fermions 
can have also a nonzero electric dipole moments.
Equations (\ref{01-TMaj0}) and (\ref{01-PTMaj0}) imply also 
that the Majorana particles (neutrinos)
cannot couple to a real photon.
The axial current of a Majorana fermion, 
$\bar{\chi}_{k}(x)\,\gamma_{\alpha}\,\gamma_5\chi_{k}(x) \neq 0$.
Correspondingly, $\chi_{k}(x)$ can have an effective 
coupling to a {\it virtual} photon 
via the {\it anapole momentum} term, which 
has the following form in momentum space:
\beq
\left ( g_{\alpha \beta}\,q^2 - q_{\alpha}\,q_{\beta}\right )\, 
\gamma_{\beta}\,\gamma_5\, F^{(k)}_{a}(q^2)\,, 
\label{anapole}
\eeq
%
where $q$ is the momentum of the virtual photon 
and  $F^{(k)}_{a}(q^2)$ is the anapole formfactor of $\chi_{k}$.
The fact that the vector current of  $\chi_{k}$ is zero 
while the axial current is nonzero has important implications 
in the calculations of the relic density of the 
lightest and stable neutralino, which is a Majorana 
particle and the dark matter candidate in 
many supersymmetric (SUSY) extensions 
of the Standard Model \cite{01-HG83}.

 In certain cases (e.g., in theories with a keV mass 
Majorana neutrino (see, e.g., \cite{01-sterile}), 
in the TeV scale type I see-saw model 
(see, e.g., \cite{01-TeVseesaw}), 
in SUSY extensions of the Standard Model)
one can have effective interactions 
involving two different massive Majorana fermions (neutrinos), 
say $\chi_1$ and $\chi_2$.
We will consider two examples. The first is an effective 
interaction with the photon field, which 
can be written as: 
\beq 
{\cal L}^{({\rm A})}_{\rm eff}(x) =  
\bar{\chi}_1(x)\, \sigma_{\alpha \beta} 
(\mu_{12} - d_{12}\gamma_5)\,\chi_2(x)\,
F^{\alpha \beta}(x) +  \hbox{h.c.} \,,
\label{01-Leff12A}
\eeq
%
where $\mu_{12}$ and $d_{12}$
are, in general, complex constants, 
$F^{\alpha \beta}(x) = \partial^{\alpha}A^{\beta}(x) - 
\partial^{\beta}A^{\alpha}(x)$,
$A^{\mu}(x)$ being the 4-vector potential 
of the photon field. Using the Majorana
conditions for $\chi_1(x)$ and $\chi_2(x)$ in the 
convention in which the phases $\xi_1 = \xi_2 = 1$,
it is not difficult to show that the constants 
$\mu_{12}$ and $d_{12}$ enter into the expression for 
${\cal L}^{({\rm A})}_{\rm eff}(x)$ in the form:
$(\mu_{12} - \mu^{\ast}_{12}) = 2i{\rm Im}(\mu_{12}) \equiv \tilde{\mu}_{12}$, 
$(d_{12} + d^{\ast}_{12}) = 2{\rm Re}(d_{12}) \equiv \tilde{d}_{12}$, 
i.e., $\tilde{\mu}_{12}$ is purely imaginary and 
$\tilde{d}_{12}$ is real
\footnote{In the case of $\chi_1(x) \equiv \chi_2(x) = \chi(x)$, 
the current $\bar{\chi}(x) \sigma_{\alpha \beta} 
(\mu_{12} - d_{12}\gamma_5)\chi_(x)$ has to be hermitian, 
which implies that $\tilde{\mu}_{12}$ should be real
while $\tilde{d}_{12}$ should be purely imaginary.
Combined with constraints on  $\tilde{\mu}_{12}$ and 
$\tilde{d}_{12}$ we have just obtained, this leads to 
 $\tilde{\mu}_{12} = \tilde{d}_{12} = 0$, which is 
consistent with 
eqs. (\ref{01-TMaj0}) and (\ref{01-PTMaj0}).}. 
In the case of CP invariance of 
${\cal L}^{({\rm A})}_{\rm eff}(x)$, the constants 
$\mu_{12}$ ($\tilde{\mu}_{12}$) and 
$d_{12}$ ($\tilde{d}_{12}$) should satisfy:
\beq
{\rm CP~invariance}:~~
\mu_{12}  =  -\,\rho_1\,\rho_2\,\mu_{12}\,,~
d_{12} = +\,\rho_1\,\rho_2\,d_{12}\,.
\label{01-CPtmutd}
\eeq
%
Thus, if  $\rho_1 = \rho_2$, i.e., if 
$\chi_1(x)$ and $\chi_2(x)$
posses the same CP-parity, 
$\mu_{12} = 0$ 
and $d_{12}$ (and $\tilde{d}_{12}$)
can be different from zero. If  $\rho_1 = -\,\rho_2$, 
i.e., if $\chi_1(x)$ and $\chi_2(x)$
have opposite CP-parities, 
$d_{12} = 0$ and $\mu_{12}$ (and $\tilde{\mu}_{12}$)
can be different from zero. 
If CP invariance does not hold, we can have both 
$\mu_{12} \neq 0$ and  $d_{12}\neq 0$
($\tilde{\mu}_{12} \neq 0$ and  $\tilde{d}_{12}\neq 0$).

 As a second example we will consider effective 
interaction of $\chi_1$ and $\chi_2$ with a vector field (current), 
which for concreteness will be assumed to be the $Z^0$-boson field 
of the Standard Model: 
\beq 
{\cal L}^{({\rm Z})}_{\rm eff}(x) =  
\bar{\chi}_1(x)\, \gamma_{\alpha} 
(v_{12} - a_{12}\gamma_5)\,\chi_2(x)\,Z^{\alpha}(x) +  \hbox{h.c.} \,,
\label{01-Leff12Z}
\eeq
%
Here $v_{12}$ and  $a_{12}$ are, in general, complex constants.
Using the Majorana conditions for  $\chi_1(x)$ and $\chi_2(x)$ with 
$\xi_1 = \xi_2 = 1$, one can easily show that 
 $v_{12}$ has to be purely imaginary, 
while $a_{12}$ has to be real 
\footnote{In the case of  $\chi_1(x) \equiv \chi_2(x) = \chi(x)$,
the hermiticity of the current $\bar{\chi}(x) \gamma_{\alpha} 
(v_{12} - a_{12}\gamma_5) \chi(x)$ implies that 
both $v_{12}$ and $a_{12}$  have to be real.
This, together with constraints on $v_{12}$ and $a_{12}$
just derived, leads to $v_{12}=0$, which is consistent 
with the result given in eq. (\ref{01-VMaj0}).}.
The requirement of CP invariance of 
${\cal L}^{({\rm Z})}_{\rm eff}(x)$, as can be shown, leads 
to ($\xi_1 = \xi_2 = 1$):
\beq
{\rm CP~invariance}:~~
v_{12} = -\,\rho_1\,\rho_2\,v_{12}\,,~~
a_{12} = +\,\rho_1\,\rho_2\,a_{12}\,.
\label{01-CPtvta}
\eeq
%
Thus, we find, similarly to the case considered above,
that if $\chi_1(x)$ and $\chi_2(x)$
posses the same CP-parity ($\rho_1 = \rho_2$), 
$v_{12} = 0$ and $a_{12}$ can be 
different from zero;
if $\chi_1(x)$ and $\chi_2(x)$ have 
opposite CP-parities ($\rho_1 = -\,\rho_2$), 
$a_{12} = 0$ while $v_{12}$ can be different from zero. 
If CP invariance does not hold, we can have both 
$v_{12} \neq 0$ and  $a_{12}\neq 0$.

 These results have important implications, in particular, 
for the phenomenology of the heavy Majorana neutrinos 
$N_k$ in the TeV scale (type I) see-saw models,
for the neutralino phenomenology in SUSY extensions of 
the Standard Model, in which the neutralinos are 
Majorana particles, and more specifically for 
the processes $e^- + e^+ \rightarrow \chi_1 + \chi_2$,
$\chi_2 \rightarrow \chi_1 + l^+ + l^-$ 
($m(\chi_2) > m(\chi_1)$), $l=e,\mu,\tau$, 
where $\chi_1$ and $\chi_2$ are, for example, 
two neutralinos of, e.g., the minimal SUSY extension of the 
Standard Model 
(see, e.g., \cite{01-eechi1chi286,01-SUSYMajCP}).

  Finally, if $\Psi(x)$ is a Dirac field 
and we define the standard propagator of $\Psi(x)$ as
\beq
 <0|T(\Psi_{\alpha}(x)\bar{\Psi}_{\beta}(y))|0> = 
S_{\alpha\beta}^{F}(x-y)~,
\label{01-DiracP} 
\eeq
%
%
\noindent one has
\beq
<0|T(\Psi_{\alpha}(x)\Psi_{\beta}(y))|0> = 0~,~~
<0|T(\bar{\Psi}_{\alpha}(x)\bar{\Psi}_{\beta}(y))|0> = 0~.
\label{01-DNSProp}
\eeq
%
\noindent In contrast, a Majorana neutrino 
field $\chi_k(x)$ has, in addition to the standard propagator
\beq
<0|T(\chi_{k\alpha}(x)\bar{\chi}_{k\beta}(y))|0> = 
S_{\alpha\beta}^{Fk}(x-y)~,
\label{01-MajSP}
\eeq
%
%
\noindent two non-trivial {\it non-standard (Majorana)} propagators
\begin{eqnarray}
<0|T(\chi_{k\alpha}(x)\chi_{k\beta}(y))|0> = 
-\xi^{*}_k S_{\alpha\delta}^{Fk}(x-y) C_{\delta\beta}~,\\[0.2cm]
<0|T(\bar{\chi}_{k\alpha}(x)\bar{\chi}_{k\beta}(y))|0> = 
\xi_k~C^{-1}_{\alpha\delta}S_{\delta\beta}^{Fk}(x-y)~.
\label{01-MajNSP}
\end{eqnarray}
%
%
\noindent This result implies that 
if $\nu_j(x)$ in eq. (\ref{eq-01:CC}) 
are massive Majorana neutrinos,
$\betabeta$-decay can proceed
by exchange of virtual neutrinos $\nu_j$ since  
$<0|T(\nu_{j\alpha}(x)\nu_{j\beta}(y))|0> \neq 0$.
The Majorana propagators play a crucial role 
in the calculation of the baryon asymmetry of 
the Universe in the leptogenesis scenario of the 
asymmetry generation (see, e.g., \cite{01-LG1,01-Branco2012}).

\subsection{Generating Dirac and Majorana Massive Neutrinos}
\label{sec-01:third} 

 The type of massive neutrinos in a given theory 
is determined by the type of the (effective) mass term 
$\mathcal{L}_{m}^{\nu}(x)$ neutrinos have,
more precisely, by the symmetries 
of $\mathcal{L}_{m}^{\nu}(x)$ and of the total 
Lagrangian $\mathcal{L}(x)$ of the theory.
A fermion mass term is any bi-linear in the fermion 
fields which is invariant under the proper 
Lorentz transformations.

   Massive Dirac neutrinos arise in theories 
in which the neutrino mass term conserves 
some additive quantum number that could be, e.g., 
the (total) lepton charge $L = L_e + L_{\mu} + L_{\tau}$,
which is conserved also by the total Lagrangian 
$\mathcal{L}(x)$ of the theory. A well known example is the 
Dirac mass term, which can arise in the minimally 
extended Standard Model to include three RH neutrino 
fields $\nu_{lR}$, $l=e,\mu,\tau$, 
as $SU(2)_{\rm L}$ singlets: 
\beq 
\mathcal{L}_{D}^{\nu}(x) =  
- ~\overline{\nu_{l'R}}(x)~M_{Dl'l}~\nu_{lL}(x) + h.c.~\,,
\label{01-Dmass1}
\eeq
%
where $M_D$ is a $3\times 3$, in general complex, matrix.
The term $\mathcal{L}_{D}^{\nu}(x)$ can be generated 
after the spontaneous breaking of the Standard Model gauge 
symmetry by an $SU(2)_{\rm L}\times U(1)_{\rm Y_w}$ invariant 
Yukawa coupling of the lepton doublet, Higgs doublet and 
the RH neutrino fields \cite{01-STP77}:
\begin{eqnarray}
\label{01-Ynu1}
{\cal L}_{\rm Y}(x) &=& 
-\,Y^{\nu}_{l' l}\,\overline{\nu_{l' R}}(x)\,H^{\dagger}(x)\,\psi_{l L}(x) 
+ \hbox{h.c.}\,,\\
 M_{D} &=&  v\,Y^{\nu}\,.
\label{01-Dmass2}
\end{eqnarray}
%
If the nondiagonal elements of $M_D$ 
are different from zero, 
$M_{Dl'l}\neq 0$, $l\neq l'=e,\mu,\tau$, 
the individual lepton charges $L_l$, $l=e,\mu,\tau$, 
will not be conserved.
Nevertheless, the total lepton charge $L$ is conserved by 
$\mathcal{L}_{D}^{\nu}(x)$. As in the case of the 
charged lepton and quark mass matrices generated 
via the spontaneous electroweak symmetry breaking 
by Yukawa type terms in the SM Lagrangian,
$M_D$ is diagonalised by a bi-unitary transformation:
$M_{D} = U^{lep}_R M^{diag}_{D} (U^{lep}_L)^{\dagger}$, 
where $U^{lep}_R$ and $U^{lep}_L$ 
are $3 \times 3$ unitary matrices. If the 
mass term in eq. (\ref{01-Dmass1}) is written 
in the basis in which the charged lepton mass 
matrix is diagonal,  $U^{lep}_L$ coincides with 
the PMNS matrix, $U^{lep}_L\equiv U_{\rm PMNS}$.
The neutrinos $\nu_j$ with definite mass $m_j > 0$
are Dirac particles: their fields 
$\nu_j(x) = (U^{lep}_L)^{\dagger}_{jl} \nu_{lL}(x) + 
(U^{lep}_R)^{\dagger}_{jl'} \nu_{l'R}(x)$
do not satisfy the Majorana condition,
$C~(\bar{\nu}_j(x))^{{\rm T}} \neq \xi_j \chi_j(x)$.
Although the scheme we are considering is 
phenomenologically viable
\footnote{It does not contain a candidate for 
a dark matter particle.}, 
it does not provide an insight of why 
the neutrino masses are much smaller than 
the charged fermion masses. The only observable 
``new physics'' is that related to the 
neutrino masses and mixing: apart from the 
neutrino masses and mixing themselves, this is the 
phenomenon of neutrino oscillations 
\cite{01-STP77}. 

  Indeed, given the fact that 
the lepton charges $L_l$, $l=e,\mu,\tau$, 
are not conserved, processes like 
$\mu^+ \rightarrow e^+ + \gamma$ decay, 
$\mu^- \rightarrow e^- + e^+ + e^-$ decay, 
$\tau^- \rightarrow e^- + \gamma$ decay, etc.
are allowed. However, the rates of these processes 
are suppressed by the factor \cite{01-STP77}
$|U_{l'j}U^{*}_{lj}m^2_j/M^2_W|^2$, $l'\neq l$, 
$M_W \cong 80$ GeV being the $W^\pm$-mass
and $l=\mu$, $l'=e$ for the 
$\mu^\pm \rightarrow e^\pm + \gamma$ decay, etc.,
and are unobservably small. For instance,
for the $\mu \rightarrow e + \gamma$ 
decay branching ratio we have \cite{01-STP77} 
(see also \cite{01-BPP77}):
\beq
BR(\mu \rightarrow e + \gamma) = \frac{3\alpha}{32\pi}\,
\left | U_{ej}\,U^{*}_{\mu j}\, \frac{m^2_j}{M^2_W} \right |^2 
\cong (2.5 - 3.9)\times 10^{-55}\,, 
\label{muegDnus}
\eeq
%
where we have used 
the best fit values of the neutrino oscillation 
parameters given in eqs.(\ref{01-Delta2131}) and 
(\ref{01-sinsq1213}) and the two values 
correspond to $\delta = \pi$ and 0. 
The current experimental upper limit 
reads \cite{01-Adam:2013}: 
$BR(\mu^+\rightarrow e^+ + \gamma) < 5.7\times 10^{-13}$.
Thus, although the predicted branching ratio 
$BR(\mu^+ \rightarrow e^+ + \gamma)\neq 0$, 
its value is roughly by 43 orders of magnitude 
smaller than the sensitivity reached in the 
experiments searching for 
the $\mu \rightarrow e + \gamma$ decay,  
which renders it unobservable in practice. 

  As was emphasised already, massive Majorana neutrinos 
appear in theories with no conserved 
additive quantum number, and more specifically, 
in which the total lepton charge $L$ is not 
conserved and changes by two units. 
In the absence of RH singlet neutrino 
fields in the theory, the flavour neutrinos 
and antineutrinos $\nu_l$ and $\bar{\nu}_l$, 
$l=e,\mu,\tau$, can have a mass term 
of the so-called Majorana type:
\beq
\mathcal{L}_{M}^{\nu}(x) =  
- ~\frac{1}{2}~\overline{\nu^c_{l'R}}(x)~M_{l'l}~\nu_{lL}(x) + h.c.~,~
\nu^c_{l'R} \equiv C~(\overline{\nu_{l'L}}(x))^{{\rm T}}\,,
\label{01-Majmasst}
\eeq
%
where $M$ is a $3\times 3$, in general complex, matrix.
In the case when all elements of $M$ are nonzero, 
$M_{l'l}\neq 0$, $l,l'=e,\mu,\tau$,
neither the individual lepton charges $L_l$ nor the total 
lepton charge $L$ are conserved:  
$L_l\neq\, const.$, $L\neq\, const.$
As it is possible to show,
owing to the fact that $\nu_{lL}(x)$ are 
fermion (anti-commuting) fields, the matrix $M$ 
has to be symmetric (see, e.g., \cite{01-BiPet87}): 
$M = M^{\rm T}$. A complex symmetric matrix is diagonalised 
by the {\it congruent transformation}:
\beq
M^{diag} = U^{\rm T} M U\,,~U-{\rm unitary}\,,
\label{01-CongrT}
\eeq
%
where $U$ is a $3\times3$ unitary matrix. 
If $\mathcal{L}_{M}^{\nu}(x)$ is written in the basis in which 
the charged lepton mass matrix is diagonal, 
$U$ coincides with the PMNS matrix: 
$U\equiv U_{\rm PMNS}$. The fields of neutrinos $\nu_j$ 
 with definite mass $m_j$ are expressed in terms of 
$\nu_{lL}(x)$ and $\nu^c_{lR}$:
\begin{eqnarray}
\mathcal{L}_{M}^{\nu}(x) &=&  
- ~\frac{1}{2}\,\overline{\nu_{j}}(x)\,m_j\,\nu_{j}(x)\,,~\\
\nu_j(x) &=& U^{\dagger}_{jl}\nu_{lL}(x) + U^{T}_{jl}\nu^c_{lR} 
= C~(\overline{\nu_{j}}(x))^{{\rm T}}\,,~~j=1,2,3\,.
\label{01-Majnus}
\end{eqnarray}
%
They satisfy the Majorana condition with $\xi_j =1$, 
as eq. (\ref{01-Majnus}) shows.

  The Majorana mass term (\ref{01-Majmasst})
for the LH flavour neutrino fields $\nu_{lL}$ 
can be generated\\ 
i) effectively after the electroweak symmetry (EWS) breaking 
in the type I see-saw models \cite{01-seesaw},\\
ii) effectively after the 
EWS breaking in the type III see-saw models \cite{01-seesawIII},\\
iii) directly as a result of the EWS breaking 
by an $SU(2)_L$ triplet Higgs field which carries two units 
of the weak hypercharge $Y_W$ and couples in an 
$SU(2)_L\times U_{Y_W}$ invariant manner 
to two lepton doublets \cite{01-seesawII} 
(the Higgs Triplet Model (HTM) sometimes 
called also ``type II see-saw model''), \\
iv) as a one loop correction to a 
Lagrangian which does not contain 
a neutrino mass term \cite{01-1loopMaj} 
(see also \cite{01-3loopMaj}), \\
iv) as a two loop correction in a theory 
where the neutrino masses are zero at tree 
and one loop levels \cite{01-2loopMaj} 
(see also \cite{01-3loopMaj}), \\
v) as a three loop correction in a theory 
in which  the neutrino masses are zero at tree,  
one loop and two loop levels \cite{01-3loopMaj}.\\
In all three types of see-saw models, for instance, 
the neutrino masses can be generated at the EWS breaking 
scale and in this case the models predict rich beyond the Standard 
Model physics at the TeV scale, some of which can 
be probed at the LHC (see, e.g., 
\cite{01-Cely:2012bz} and further).   
We will consider briefly below the neutrino mass generation in 
the type I see-saw and the Higgs triplet models.

 In a theory in which the $SU(2)_{\rm L}$ singlet 
RH neutrino fields $\nu_{lR}$, 
$l=e,\mu,\tau$, are present 
\footnote{We consider in the present article 
the case of three RH sterile neutrinos, 
but schemes with less than 3 and more than 
3 sterile neutrinos are also discussed in the 
literature, see, e.g., \cite{01-Thnu07,01-sterile}.
},
the most general neutrino mass Lagrangian  
contains the Dirac mass term (\ref{01-Dmass1}), 
the Majorana mass term for the 
LH flavour neutrino fields (\ref{01-Majmasst})
and a Majorana mass term 
for the RH neutrino fields $\nu_{lR}(x)$ 
\cite{01-BiPont76}: 
\beq 
\mathcal{L}_{D+M}^{\nu}(x) =  
-~\overline{\nu_{l'R}}(x)~M_{Dl'l}~\nu_{lL}(x) 
-~\frac{1}{2}~\overline{\nu^c_{l'R}}(x)~M_{Ll'l}~\nu_{lL}(x)
-~\frac{1}{2}~\overline{\nu^c_{l'L}}(x)~M_{Rl'l}~\nu_{lR}(x)
+ h.c.~\,,
\label{01-DMmasst}
\eeq
where $\nu^c_{l'L}\equiv C~(\overline{\nu_{l'R}}(x))^{{\rm T}}$
and $M_D$, $M_L$ and $M_R$ are  $3\times 3$, in general complex, 
matrices.
By a simple rearrangement of the neutrino fields 
this mass term can be cast in the form of a 
Majorana mass term which is then 
diagonalised with the help of the congruent 
transformation \cite{01-BiPet87}.
In this case there are six Majorana mass 
eigenstate neutrinos, i.e., the flavour neutrino 
fields $\nu_{lL}(x)$ are linear combinations of 
the fields of six Majorana neutrinos 
with definite mass. The neutrino mixing 
matrix in eq. (\ref{eq-01:CC}) is a $3\times 6$ 
block of a $6\times 6$ unitary matrix.   

  The Dirac-Majorana mass term is at the basis 
of the type I see-saw mechanism of generation 
of the neutrino masses and 
appears in many Grand Unified Theories (GUTs)
(see, e.g., \cite{01-BiPet87} for further details). 
In the see-saw models, 
some of the six massive Majorana 
neutrinos typically are too heavy to be produced in
the weak processes in which the initial states of the 
flavour neutrinos and antineutrinos
$\nu_\ell$ and $\bar{\nu}_\ell$, 
used in the neutrino oscillation experiments, 
are being formed.
As a consequence, the states of 
$\nu_\ell$ and $\bar{\nu}_\ell$ will be coherent 
superpositions only of the states of the light massive 
neutrinos $\nu_j$, and  the elements of  
the neutrino mixing matrix $U_{\rm PMNS}$, which
are determined in experiments studying 
the oscillations of 
$\nu_\ell$ and $\bar{\nu}_\ell$, 
will exhibit deviations from unitarity.
These deviations can be relatively large 
and can have observable effects 
in the TeV scale see-saw models, 
in which the heavy Majorana neutrinos have 
masses in the $\sim (100 - 1000)$ GeV range 
(see, e.g., \cite{01-Antusch:2006vwa}).

If after the diagonalisation of $\mathcal{L}_{D+M}^{\nu}(x)$
more than three neutrinos will turn out to be light, i.e., 
to have masses $\sim 1$ eV or smaller, 
active-sterile neutrino oscillations can take place  
(see, e.g., \cite{01-sterile,01-BiPet87}):
a LH (RH) flavour neutrino $\nu_{lL}$ 
(antineutrino $\bar{\nu}_{lR}$)  
can undergo transitions into LH sterile 
antineutrino(s) $\bar{\nu}_{l'L} \equiv \overline{\nu^s}_{l'L}$ 
(RH sterile neutrino(s) $\nu_{l'R} \equiv \nu^s_{l'R}$). 
As a consequence of this type of oscillations,
one would observe a ``disappearance'' of, e.g., 
$\nu_e$ and/or $\nu_{\mu}$ 
($\bar{\nu}_e$ and/or $\bar{\nu}_{\mu}$)  
on the way from the source to the detector.

 We would like to discuss next 
the implications of CP invariance for the 
neutrino Majorana mass matrix, eq. (\ref{01-Majmasst}).
In the convention we have used to derive 
eqs. (\ref{01-YCPinv1}) and (\ref{01-RCPinv1}),
in which the unphysical phase factor in the 
CP transformation of the lepton doublet field 
$\Psi_{lL}(x)$, and thus of $\nu_{lL}(x)$, $\eta^l = i$,  
the requirement of CP invariance leads to 
the reality condition for $M$:
\beq 
{\rm CP-invariance:}~~~ M^{*} = M\,.
\label{01-MCPinv1}
\eeq
%
Thus, $M$ is real and symmetric 
and therefore is diagonalised by 
an orthogonal transformation, i.e., 
if CP invariance holds, the matrix 
$U$ in eq. (\ref{01-CongrT}) is an  
orthogonal matrix. The nonzero eigenvalues of 
a real symmetric matrix can be 
positive or negative
\footnote{The absolute value of the difference between 
the number of positive and number of negative eigenvalues 
of a real symmetric matrix $A$ is an invariant of the matrix 
with respect to transformations $A' = PAP^T$,
where $P$ is a real matrix which has an inverse.
}. 
Consequently, $M^{diag}$ in eq. (\ref{01-CongrT}) in general has the form:
\beq
M^{diag} = (m'_1,m'_2,m'_3)\,,~~m'_j = \rho_j\,m_j\,,~m_j > 0\,,~ 
\rho_j = \pm 1\,.
\label{01-Mdiag1}
\eeq
%
Let us denote the neutrino field 
which has a mass $m'_j \neq 0$ by $\nu'_j(x)$.
According to eq.  (\ref{01-Majnus}), the field 
$\nu'_j(x)$ satisfies the Majorana condition:
$\nu'_j(x)= C~(\overline{\nu'_{j}}(x))^{{\rm T}}$.
One can work with the fields $\nu'_j(x)$ remembering that 
some of them have a negative mass. 
It is not difficult to show that 
the CP-parity of the fields 
$\nu'_j(x)$ is $\eta_{CP}(\nu'_j) = i$, 
$j=1,2,3$. The physical meaning 
of the signs of the masses $m'_j \neq 0$
of the Majorana neutrinos 
becomes clear if we change to a ``basis'' 
of neutrino fields $\nu_j(x)$ 
which have positive masses $m_j > 0$. 
This can be done, e.g., by introducing the 
fields \cite{01-BiPet87}:
\beq
\nu'_j(x) = (-\,\gamma_5)^{\frac{1}{2}(1-\rho_j)}\nu_j(x)\,:~
\nu'_j(x) = \nu_j(x)\,~{\rm if}~\rho_j = 1\,;~
\nu'_j(x) = -\,\gamma_5\,\nu_j(x)\,~{\rm if}~\rho_j = -\, 1\,.
\label{01-nuj}
\eeq
%
As it is not difficult to show, 
if $\nu'_j(x)$ has a mass $m'_j < 0$, 
CP-parity  $\eta_{CP}(\nu'_j) = i$ and satisfies the 
Majorana condition 
$C~(\overline{\nu'_{j}}(x))^{{\rm T}} = \rho_j\nu'_j(x)$,
the field $\nu_j(x)$ posses a mass $m_j > 0$, 
CP-parity  $\eta_{CP}(\nu_j) = i\rho_j$, 
and satisfy the Majorana condition 
$C~(\overline{\nu_{j}}(x))^{{\rm T}} = \rho_j\nu_j(x)$:
\beq
\nu_j\,:~~m_j > 0\,,~~\eta_{CP}(\nu_j) = i\rho_j\,,~~
C~(\overline{\nu_{j}}(x))^{{\rm T}} = \rho_j\nu_j(x)\,.
\label{01-nuj2}
\eeq
%
Thus, in the case of CP invariance, the signs 
of the nonzero eigenvalues of the neutrino 
Majorana mass matrix determine the CP-parities of 
the corresponding positive mass Majorana 
(mass eigenstate) neutrinos 
\footnote{
For further discussion of the properties of massive 
Majorana neutrinos (fermions) and their couplings
see, e.g., \cite{01-BiPet87}.
}.

\subsection{A Brief Historical Detour}

 It is interesting to note that B. Pontecorvo in his  
second seminal article on neutrino oscillations \cite{01-BPont57}, 
which was published 
in 1958 when only one type of neutrino and antineutrino was known, 
assumed that the state of the neutrino $\nu$, emitted in weak interaction 
processes is a linear superposition of the states 
of two Majorana neutrinos $\nu^M_1$ and $\nu^M_2$ which have 
different masses, $m_1 \neq m_2$, opposite CP-parities, 
$\eta_{CP}(\nu^M_1) = - \eta_{CP}(\nu^M_2)$ and are maximally mixed,
while the state of the corresponding antineutrino $\overline{\nu}$ 
is just the orthogonal superposition of the states of  
$\nu^M_1$ and $\nu^M_2$: 
\begin{eqnarray}
|\nu> &=& \frac{|\nu^M_1 > +\, |\nu^M_2 >}{\sqrt{2}}\,,\\
|\overline{\nu}> &=& \frac{|\nu^M_1 > - \,|\nu^M_2 >}{\sqrt{2}}\,.
\end{eqnarray}
%
Thus, the oscillations are between the neutrino $\nu$ and
and the antineutrino $\overline{\nu}$,  in full analogy with 
the $K^0 - \overline{K}^0$ oscillations. 
From contemporary point of view, 
B. Pontecorvo proposed active - sterile neutrino oscillations 
with maximal mixing and massive Majorana neutrinos.
To our knowledge, the 1958 article quoted in \cite{01-BPont57} 
was also the first in which fermion mixing in the weak 
interaction Lagrangian was introduced. 

  The article of Z. Maki, M. Nakagawa and S. Sakata \cite{01-MNS62}
was inspired, in part, by the discovery of the second type 
of neutrino - the muon neutrino, in 1962 at Brookhaven.
These authors considered a composite model of elementary 
particles in which the electron and muon neutrino states 
are superpositions of the states of {\it composite Dirac neutrinos} 
$\nu^D_1$ and $\nu^D_2$ which have different masses, 
 $m^D_1 \neq m^D_2$:
\begin{eqnarray}
|\nu_e> &=& |\nu^D_1 >\, \cos\theta_c + |\nu^D_2 >\,\sin\theta_c\,,\\
|\nu_{\mu}> &=& -\,|\nu^D_1 >\, \sin\theta_c + |\nu^D_2 >\,\cos\theta_c\,,
\end{eqnarray}
%
where $\theta_c$ is the neutrino mixing angle. 
The model proposed in \cite{01-MNS62} has lepton-hadron 
symmetry built in and as consequence of this symmetry the 
neutrino mixing angle coincides with what we call 
today the Cabibbo angle
\footnote{The article by  Z. Maki, M. Nakagawa and 
S. Sakata \cite{01-MNS62} appeared before the 
article by N. Cabibbo \cite{01-Cabibbo63} in which the 
``Cabibbo angle'' $\theta_c$ was introduced and  
the hadron phenomenology related to this angle 
was discussed, but after the article by M. Gell-Mann and 
M. Levy \cite{01-GellMannLevy60}
in which $\theta_c$ was also introduced 
(by the way, in a footnote).
} 
$\theta_c \cong 0.22$.
The authors of \cite{01-MNS62} discuss 
the possibility of $\nu_{\mu} - \nu_e$ 
oscillations, which they called ``virtual transmutations''.

   In an article \cite{01-KMTY62} by  Y. Katayama, K. Matsumoto, 
S. Tanaka and E, Yamada, 
published in 1962 somewhat earlier than \cite{01-MNS62},
the authors also introduce two-neutrino mixing. 
However, this is done purely for model construction 
purposes and does not have any physical consequences 
since the neutrinos in the model constructed in 
\cite{01-KMTY62} are massless particles.

   In 1967 B. Pontecorvo independently 
considered the possibility of $\nu_e \leftrightarrow \nu_{\mu}$ oscillations 
in the article \cite{01-BPont67}, in which the notion of a 
``sterile'' or ``inert'' neutrino was introduced.
Later in 1969, V. Gribov and B. Pontecorvo \cite{01-VGBPont69}
introduced for the first time a Majorana mass term for the 
LH flavour neutrinos $\nu_e$ and $\nu_{\mu}$, 
the diagonalisation of which lead to two Majorana 
neutrinos $\nu^M_{1,2}$  with definite but different
masses $m_{1,2}$,  $m_{1} \neq m_{2}$, and 
two-neutrino mixing with an arbitrary 
mixing angle $\theta$:
\begin{eqnarray}
|\nu_e> &=& |\nu^M_1 >\, \cos\theta + |\nu^M_2 >\,\sin\theta\,,\\
|\nu_{\mu}> &=& -\,|\nu^M_1 >\, \sin\theta + |\nu^M_2 >\,\cos\theta\,.
\end{eqnarray}
%
This was the first modern treatment of the problem of 
neutrino mixing which anticipated the way this problem 
is addressed in gauge theories of electroweak interactions 
and in Grand Unified Theories (GUT's).
In the same article for the first time the 
analytic expression for the probability  
of $\nu_e \leftrightarrow \nu_{\mu}$ oscillations 
was also derived.

\subsection{Models of Neutrino Mass Generation: Two Examples}  

{\bf Type I See-Saw Model.} A natural explanation of the 
smallness of neutrino masses is provided by 
the type I see-saw mechanism of neutrino 
mass generation \cite{01-seesaw}. An integral part 
of this rather simple mechanism 
are the RH neutrinos $\nu_{lR}$
(RH neutrino fields $\nu_{lR}(x)$). The latter are
assumed to possess a Majorana mass term 
${\cal L}^{\rm N}_{\rm M}(x)$ as well as
Yukawa type coupling ${\cal L}_{\rm Y}(x)$ with
the Standard Model lepton and Higgs doublets, 
$\psi_{lL}(x)$ and $H(x)$, given in eq. (\ref{01-Ynu1}). 
In the basis in which the Majorana mass
matrix of RH neutrinos is diagonal, we have:
\beq
\label{01-Ynu2}
{\cal L}_{\rm Y,M}(x) \equiv  {\cal L}_{\rm Y}(x) + 
{\cal L}^{\rm N}_{\rm M}(x) =
-\,\left ( \lambda_{k l}\,\overline{N_{kR}}(x)\, H^{\dagger}(x)\,\psi_{l L}(x)
+ \hbox{h.c.}\right )
-\,\frac{1}{2}\,M_{k}\,\overline{N_k}(x)\, N_k(x)\,,
\eeq
%
\noindent
where we have combined the expressions given in 
eqs. (\ref{01-Ynu}) and  (\ref{01-MN}).
When the electroweak
symmetry is broken spontaneously,
the neutrino Yukawa coupling generates a Dirac mass term:
$m^{D}_{kl}\,\overline{N_{kR}}(x)\,\nu_{l L}(x) + \hbox{h.c.}$,
with $m^{D} = v\lambda$, $v = 174$ GeV being the Higgs doublet v.e.v.
In the case when the elements of
$m^{D}$ are much smaller than $M_k$,
$|m^{D}_{il}|\ll M_k$, $i,k=1,2,3$, $l=e,\mu,\tau$,
the interplay between the
Dirac mass term and the mass term
of the heavy (RH) Majorana neutrinos
$N_k$ generates an effective Majorana mass
(term) for the LH flavour neutrinos \cite{01-seesaw}:
\beq
M_{l'l} \cong -\, (m^{D})^T_{l'k}\,M^{-1}_{k}\,m^{D}_{kl} = 
 -\,v^2\,  (\lambda)^T_{l'k}\,M^{-1}_{k}\,\lambda_{kl}\,.
\label{01-MajMSeeS}
\eeq
%
In grand unified theories, $m^{D}$ is typically of the
order of the charged fermion masses. In $SO(10)$ theories,
for instance, $m^{D}$ coincides with the up-quark mass matrix.
Taking indicatively 
$M\sim 0.1$ eV, $m^{D}\sim 100$ GeV, one obtains 
$M_k\sim M_N \sim 10^{14}$ GeV, which is close
to the scale of unification of the electroweak and
strong interactions, $M_{GUT}\cong 2\times 10^{16}$ GeV.
In GUT theories with RH neutrinos one finds
that indeed the heavy Majorana neutrinos
$N_k$ naturally obtain masses which are
by few to several orders of magnitude smaller
than $M_{GUT}$ 
(see, e.g., the second and third 
articles quoted in ref. \cite{01-seesaw}). 
Thus, the enormous
disparity between the neutrino and
charged fermion masses
is explained in this approach by
the huge difference between
effectively the electroweak
symmetry breaking scale and $M_{GUT}$.

 An additional attractive feature of the
see-saw scenario under discussion is that
the generation and smallness of neutrino
masses is related via the leptogenesis mechanism \cite{01-LG} 
(see also, e.g., \cite{01-LG1,01-FLG1,01-FLG2,01-Branco2012}) 
to the generation of the baryon asymmetry of the Universe.
Indeed, the
Yukawa coupling in (\ref{01-Ynu2}),
in general, is not CP conserving.
Due to this CP-nonconserving
coupling the heavy Majorana neutrinos
undergo, e.g., the decays
$N_j \rightarrow l^{+} + H^{(-)}$,
$N_j \rightarrow l^{-} + H^{(+)}$,
which have different rates:
$\Gamma(N_j \rightarrow l^{+} + H^{(-)}) \neq
\Gamma(N_j \rightarrow l^{-} + H^{(+)})$.
When these decays occur in the Early Universe
at temperatures somewhat below the mass of, say,
$N_1$, so that the latter are out of equilibrium with
the rest of the particles present at that epoch,
CP violating asymmetries in the
individual lepton charges $L_l$ and in the
total lepton charge $L$ of the Universe
are generated. These lepton asymmetries are
converted into a baryon asymmetry by
$(B-L)$ conserving, but $(B+L)$ violating,
sphaleron processes, which exist
in the Standard Model and are effective at
temperatures $T \sim (100 -  10^{12})$ GeV 
\cite{01-HarvTurn90}.
If the heavy neutrinos $N_j$
have hierarchical spectrum, $M_1 \ll M_2 \ll M_3$,
the observed baryon asymmetry can be reproduced
provided the mass of the lightest one satisfies 
$M_1 \gsim 10^{9}$ GeV \cite{01-DavIb02} 
~\footnote{In specific type I see-saw models this 
bound can be lower by a few orders of 
magnitude, see, e.g., \cite{01-Raidal:2004vt}.
}. 
Thus, in this scenario, the neutrino masses and mixing and
the baryon asymmetry have the same origin -
the neutrino Yukawa couplings and
the existence of (at least two)
heavy Majorana neutrinos.
Moreover, quantitative
studies based on advances
in leptogenesis theory \cite{01-FLG1,01-FLG2}, 
in which the importance of the flavour effects 
in the generation of the baryon asymmetry
was understood, have shown that the Dirac and/or Majorana
phases in the neutrino mixing matrix $U$
can provide the CP violation, necessary in leptogenesis
for the generation of the observed baryon asymmetry of the 
Universe \cite{01-PPRio106}.
This implies, in particular,
that if the CP symmetry is
established not to hold in
the lepton sector due to the PMNS matrix $U$,
at least some fraction (if not all)
of the observed baryon asymmetry
might be due to the Dirac  and/or Majorana
CP violation present in the neutrino mixing.

 In the see-saw scenario considered, 
the scale at which the new physics 
manifests itself, which is set by the scale of 
masses of the RH neutrinos, can, in principle,
have an arbitrary large value, up to the GUT scale of 
$2\times 10^{16}$ GeV and even beyond, up to 
the Planck mass. An interesting possibility, 
which can also be theoretically well motivated 
(see, e.g., \cite{01-Shaposhnikov:2006nn}), 
is to have the new physics at the TeV scale, 
i.e., $M_k \sim(100 -1000)$ GeV. 
Low scale see-saw scenarios usually predict a 
rich phenomenology at the TeV scale
and are constrained by different sets of data,
such as, e.g., the data on neutrino oscillations, 
from EW precision tests and on the 
lepton flavour violating (LFV) processes 
$\mu \rightarrow e \gamma$, $\mu \rightarrow 3e$, 
$\mu^- - e^-$ conversion in nuclei.
In the case of the TeV scale type I see-saw 
scenario of interest, the flavour structure 
of the couplings of the heavy Majorana neutrinos $N_k$   
to the charged leptons and the $W^{\pm}$ bosons, and 
to the LH flavour neutrinos $\nu_{lL}$ and the $Z^0$ boson,
are essentially determined by the requirement of 
reproducing the data on the neutrino
oscillation parameters \cite{01-TeVseesaw}.
All present experimental constraints on 
this scenario still allow 
i) for the  predicted rates of the 
$\mu\to e+ \gamma$ decay, $\mu\to 3e$ decay and $\mu-e$ 
conversion in the nuclei to be \cite{01-TeVseesaw1} within 
the sensitivity range of the currently 
running MEG experiment on $\mu\to e+ \gamma$ decay 
\cite{01-Adam:2011ch} planned to probe values of
$BR(\mu^+\rightarrow e^+ + \gamma) \gtap 10^{-13}$,
and of the future planned experiments on 
$\mu\to 3e$ decay and $\mu-e$ conversion
\cite{01-muto3eNext,01-comet,01-mu2e,01-PRIME,01-projectX},
ii) for an enhancement of the rate of 
neutrinoless double beta ($\betabeta$-) decay  
\cite{01-TeVseesaw}, which thus can be in the
range of sensitivity
of the  $\betabeta$-decay experiments which are taking data 
or are under preparation (see, e.g., \cite{01-bb0nuExp2})
even when the light Majorana neutrinos possess
a normal hierarchical mass spectrum (see further),
and 
iii) for the possibility of an exotic 
Higgs decay channel into a light 
neutrino and a heavy Majorana 
neutrino with a sizable branching ratio, 
which can lead to observable effects 
at the LHC \cite{01-Cely:2012bz}
(for further details concerning the low energy 
phenomenology of the TeV scale type I see-saw model 
see, e.g., \cite{01-TeVseesaw,01-Shaposhnikov:2006nn,01-TeVseesaw1}).

  Let us add that the role of the experiments 
searching for lepton flavour violation to test and 
possibly constrain low scale see-saw models, 
and more generally, extensions of the Standard Model 
predicting ``new'' (lepton flavour violating) physics 
at the TeV scale, will be
significantly strengthened in the next years.
Searches for $\mu-e$ conversion at the planned COMET experiment
at KEK~\cite{01-comet} and Mu2e experiment
at Fermilab~\cite{01-mu2e} aim to reach sensitivity to 
conversion rates 
$\rm{CR}(\mu\, {\rm Al} \to e\, {\rm Al})\approx  10^{-16}$,
while, in the longer run, the PRISM/PRIME experiment 
in KEK~\cite{01-PRIME}
and the project-X experiment in Fermilab~\cite{01-projectX}
are being designed to probe values of the $\mu-e$ conversion rate
on ${\rm Ti}$, which are by 2 orders of magnitude smaller,
$\rm{CR}(\mu\, {\rm Ti} \to e\, {\rm Ti})\approx 10^{-18}$~
\cite{01-PRIME}. The current upper limit on the $\mu-e$ conversion rate
is $\rm{CR}(\mu\, {\rm Al} \to e\, {\rm Al}) < 4.3\times 10^{-12}$ 
\cite{01-Dohmen:1993mp}.
There are also plans to perform a new search for the
$\mu^+\rightarrow e^+ e^- e^+$ decay \cite{01-muto3eNext},
which will probe values of the corresponding
branching ratio down to
${\rm BR}(\mu^+\rightarrow e^+ e^- e^+) \approx 10^{-15}$,
i.e., by 3 orders of magnitude smaller than
the best current upper limit \cite{01-Bellgardt:1987du}.
Furthermore, searches for tau lepton flavour violation
at superB factories aim to reach a sensitivity to
${\rm BR}(\tau\rightarrow (\mu,e)\gamma)\approx 10^{-9}$
(see, e.g., \cite{01-Akeroyd:2004mj}).\\

{\bf The Higgs Triplet Model (HTM).}
In its minimal formulation this model includes one additional
$SU(2)_{L}$ triplet Higgs field  $\Delta$, which has
weak hypercharge $Y_W=2$ \cite{01-seesawII}:
\begin{equation}
	\Delta\;=\;\left(
			\begin{array}{cc}
				\Delta^{+}/\sqrt{2} & \Delta^{++} \\
				\Delta^{0}	& -\Delta^{+}/\sqrt{2}
			\end{array}\right)\,.
\end{equation}
%
The Lagrangian of the Higgs triplet model
which is sometimes called also the ``type II see-saw model'',
reads 
\footnote{We do not give here, for simplicity,
all the quadratic and quartic
terms present in the scalar potential
(see, $e.g.$, 
\cite{01-HTMPheno2012}).
}:
\begin{eqnarray}
	\mathcal{L}_{\rm HTM} &=& 
-\, M_{\Delta}^{2}\,{\rm Tr}\left(\Delta^{\dagger}\Delta\right)
-\,\left(h_{\ell\ell^{\prime}}\,
\overline{\psi^{C}}_{\ell L}\,i\tau_{2}\,\Delta\,\psi_{\ell^{\prime}L}\,
 +\,\mu_{\Delta}\, H^{\dagger}\,\Delta^{\dagger}\,i\tau_{2}\,H^{*}\,
+\,{\rm h.c.}\right)\,,
\label{LtypeII}
\end{eqnarray}
%
where $\overline{\psi^{C}}_{\ell L}
\equiv ( -\,\nu^T_{\ell L}C^{-1}~~-\,\ell^T_{L}C^{-1})$,
$C$ being the charge conjugation matrix,
$H$ is the SM  Higgs doublet and
$\mu_{\Delta}$ is a real parameter characterising the
soft explicit breaking of the total lepton charge
conservation. We will discuss briefly the 
low energy version of HTM,
where the new physics scale $M_{\Delta}$
associated with the mass of $\Delta$ takes values
$100~{\rm GeV}\lesssim M_{\Delta} \lesssim 1~{\rm TeV}$,
which, in principle, can be probed by LHC
(see \cite{01-colliders,01-HTMPheno2012} 
and references quoted therein).

 The flavour structure of the Yukawa coupling matrix $h$ and
the size of the lepton charge soft breaking parameter $\mu_{\Delta}$
are related to the light neutrino Majorana mass matrix 
$M^{\nu}$, which is generated when the neutral component of $\Delta$
develops a ``small'' vev  $v_{\Delta} \propto \mu_{\Delta}$ .
Indeed, setting $\Delta^{0} = v_{\Delta}$ and 
$H^T  = (v~~0)^{T}$ with $v\simeq 174$ GeV,
from  Lagrangian (\ref{LtypeII}) one obtains:
\begin{equation}
M^{\nu}_{\ell\ell^{\prime}} \simeq\;2\,h_{\ell\ell^{\prime}}\,
v_{\Delta}\;.
\label{01-mnuII}
\end{equation}
%
The matrix of Yukawa couplings $h_{\ell\ell^\prime}$ is directly
related to the PMNS neutrino mixing matrix $U_{\rm PMNS}\equiv U$,
which is unitary in this case:
\begin{equation}
h_{\ell\ell^\prime}\;\equiv\; \frac{1}{2v_\Delta}\left(U^*\,
{\rm diag}(m_1,m_2,m_3)\,U^\dagger\right)_{\ell\ell^\prime}\,,~~m_j\geq 0\,.
\label{01-hU}
\end{equation}
%
An upper limit on $v_\Delta$ can be obtained from
considering its effect on the parameter $\rho=M^2_W/M_Z^2\cos^2\theta_W$.
In the SM, $\rho=1$ at tree-level, while in the HTM one has
\begin{equation}
\rho\equiv 1+\delta\rho={1+2x^2\over 1+4x^2}\,,~~~x \equiv v_\Delta/v.
\label{deltarho}
\end{equation}
%
The measurement $\rho\approx 1$ leads to the bound
$v_\Delta/v\lesssim 0.03$, or  $v_\Delta<5$~GeV
(see, e.g., \cite{01-HTMPheno2012}).

   For $M_{\Delta} \sim (100 - 1000)$ GeV, 
the model predicts a plethora of beyond the SM 
physics phenomena (see, e.g.,  
\cite{01-HTMPheno2012,01-Akeroyd:2009nu,01-HTMLHC}), 
most of which can be probed 
at the LHC and in the experiments on charged 
lepton flavour violation, if the Higgs triplet 
vacuum expectation value $v_{\Delta}$
is relatively small, roughly  
$v_{\Delta} \sim (1 - 100)$ eV. As can be shown 
(see, e.g., \cite{01-HTMPheno2012}),  
the parameters $v_{\Delta}$ and  $\mu_{\Delta}$ are related:
for $M_{\Delta} \sim v = 174$ GeV we have
$v_{\Delta} \cong \mu_{\Delta}$, while if
$M^2_{\Delta} >> v^2$, then
 $v_{\Delta} \cong \mu_{\Delta} v^2/(2M^2_{\Delta})$.
Thus, a relatively small value of $v_{\Delta}$
in the TeV scale HTM implies that $\mu_{\Delta}$ 
has also to be small, and {\it vice verse}. 
A nonzero but relatively small value 
of  $\mu_{\Delta}$ can be generated, e.g.,  
at higher orders in perturbation theory~\cite{01-Chun:2003ej}. 
The smallness of the neutrino masses is therefore 
related to the smallness of the vacuum expectation value 
$v_{\Delta}$, which in turn is related to the smallness of the 
parameter $\mu_{\Delta}$.

 Under the conditions specified 
above one can have testable predictions
of the model in low energy experiments, and in particular,
in the ongoing MEG and the planned future experiments 
on the lepton flavour violating processes 
 $\mu\rightarrow e \gamma$, $\mu\rightarrow 3e$ and
 $\mu + \mathcal{N}\to e + \mathcal{N}$ 
(see, e.g., \cite{01-TeVseesaw1}).
The HTM has also an extended Higgs sector including neutral, 
singly charged and doubly charged Higgs particles.
The physical singly-charged Higgs scalar field (particle)
practically coincides with the triplet scalar
field $\Delta^{+}$, the admixture of the 
doublet charged scalar field being suppressed by the
factor $v_{\Delta}/v$. The singly- and doubly-  charged
Higgs scalars $\Delta^{+}$ and $\Delta^{++}$ have,
in general, different masses \cite{01-Chun:2003ej}:
$m_{\Delta^{+}}\neq m_{\Delta^{++}}$.
Both cases $m_{\Delta^{+}} >  m_{\Delta^{++}}$ and
$m_{\Delta^{+}} < m_{\Delta^{++}}$ are possible.
The TeV scale HTM predicts the existence of rich new 
physics at LHC as well, associated with the presence of the 
singly and doubly charged Higgs particles  
$\Delta^{+}$ and $\Delta^{++}$ in the theory 
(see, e.g., \cite{01-HTMPheno2012,01-HTMLHC}).

\section{Determining the Nature of Massive Neutrinos}

The Majorana nature of massive neutrinos 
typically manifests itself in the existence 
of processes in which the total lepton
charge $L$ changes by  
two units:
$K^+\rightarrow \pi^- + \mu^+ + \mu^+$,
$\mu^- + (A,Z)\rightarrow \mu^+ + (A,Z-2)$, etc.
Extensive studies have shown that
the only feasible experiments 
having the potential of establishing the
Majorana nature of massive neutrinos 
at present are the
$\betabeta$-decay experiments searching 
for the process $(A,Z) \rightarrow (A,Z+2) + e^- + e^-$
(for reviews see, e.g., 
\cite{01-BiPet87,01-bb0nuExp2,01-WRodej10}).
The observation of \betabeta-decay
and the measurement of the corresponding 
half-life with sufficient accuracy,
would not only be a proof that the total 
lepton charge is not conserved, 
but might provide also 
information i) on the type of neutrino 
mass spectrum \cite{01-PPSNO2bb},
and ii) on the absolute scale of neutrino 
masses (see, e.g. \cite{01-PPW}).

 The observation of $(\beta\beta)_{0\nu}$-decay
and the measurement of the corresponding
half-life with sufficient accuracy, 
combined with data on the absolute 
neutrino mass scale might provide 
also information on the Majorana phases in $U$ 
\cite{01-BGKP96,01-BPP1,01-PPW,01-PPR1,01-PPSchw05}.
If the neutrino mass spectrum is 
inverted hierarchical or quasi-degenerate, 
for instance, it would be possible to get 
information about the phase $\alpha_{21}$.
However, establishing even in this case  
that  $\alpha_{21}$ has a CP violating value
would be a remarkably challenging problem
\cite{01-PPSchw05} (see also \cite{01-BargerCP}).
Determining experimentally
the values of both the Majorana phases 
$\alpha_{21}$ and $\alpha_{31}$ is an 
exceptionally difficult problem. 
It requires the knowledge of the 
type of neutrino mass spectrum 
and high precision determination of both 
the absolute neutrino mass scale 
and of the  $(\beta\beta)_{0\nu}$-decay
effective Majorana mass, $\meff$ 
(see, e.g., \cite{01-BPP1,01-PPR1,01-PPSchw05}).

%
\subsection{Majorana Neutrinos and $\betabeta-$Decay}
%
%
  Under the assumptions of 3-$\nu$ mixing,
for which we have compelling evidence, 
of massive neutrinos $\nu_j$ being
Majorana particles, and
of \betabeta-decay generated
{\it only by the (V-A) charged current 
weak interaction via the exchange of the three
Majorana neutrinos  $\nu_j$} having masses
$m_j \ltap$ few MeV,
the $\betabeta$-decay amplitude  
of interest 
has the form (see, e.g. \cite{01-BiPet87,01-BPP1,01-WRodej10}): 
$A\betabeta \cong \mefff~M$, where 
$M$ is the corresponding 
nuclear matrix element (NME) which does not 
depend on the neutrino mixing parameters, and
\begin{equation}
\meff = \left| m_1\, |U_{\mathrm{e} 1}|^2 
+ m_2\, |U_{\mathrm{e} 2}|^2\, e^{i\alpha_{21}}   
 + m_3\,|U_{\mathrm{e} 3}|^2\, e^{i(\alpha_{31}-2 \delta)} \right|\,,
\label{effmass2}
\end{equation}

\noindent is the effective Majorana 
mass in \betabeta-decay,
$|U_{\mathrm{e}1}| = c_{12}c_{13}$,
$|U_{\mathrm{e}2}| = s_{12}c_{13}$, 
$|U_{\mathrm{e}3}| = s_{13}$.
In the case of CP-invariance 
one has $2\delta = 0~{\rm or}~2\pi$ and
\cite{01-LW81,01-JBernaPascMajP83},
\begin{equation}
\eta_{21} \equiv e^{i\alpha_{21}} = \pm 1\,,
\eta_{31}\equiv e^{i\alpha_{31}} = \pm 1\,, 
\label{01-eta2131}
\end{equation}
%
\noindent  $\eta_{21(31)}$ being the 
relative CP-parity of the Majorana neutrinos 
$\nu_{2(3)}$ and $\nu_1$. 

   It proves convenient to express   
\cite{01-SPAS94} the three neutrino masses
in terms of $\Delta m^2_{21}$ and $\Delta m^2_{31(32)}$, measured 
in neutrino oscillation experiments,
and the absolute neutrino  mass scale
 determined by ${\rm min}(m_j)$  
\footnote{For a detailed discussion of 
the relevant formalism 
see, e.g. \cite{01-BiPet87,01-BPP1,01-WRodej10}.}.
In both cases of 
neutrino mass spectrum 
with normal and inverted ordering one has
(in the convention we use):
$\Delta m_{21}^2 > 0$, 
$m_2=(m_1^2 + \Delta m^2_{21})^{\frac{1}{2}}$.
For normal ordering, 
$\Delta m^2_{31} > 0$
and $m_3= (m_1^2 + \Delta m^2_{31})^{\frac{1}{2}}$,
while if the spectrum is with inverted ordering,
${\rm min}(m_j) = m_3$, $\Delta m_{32}^2 < 0$ and 
$m_1 = (m_3^2 + \Delta m^2_{23} - \Delta m^2_{21})^{\frac{1}{2}}$. 
Thus, given $\Delta m^2_{21}$, $\Delta m^2_{31(32)}$, $\theta_{12}$ 
and $\theta_{13}$, $\meff$ depends 
on $min(m_j)$, Majorana phases 
$\alpha_{21}$, $\alpha_{31}$ and 
the type of $\nu$-mass spectrum.

\begin{figure}[t] 
\includegraphics[width=16.0cm,height=12.0cm,angle=0]{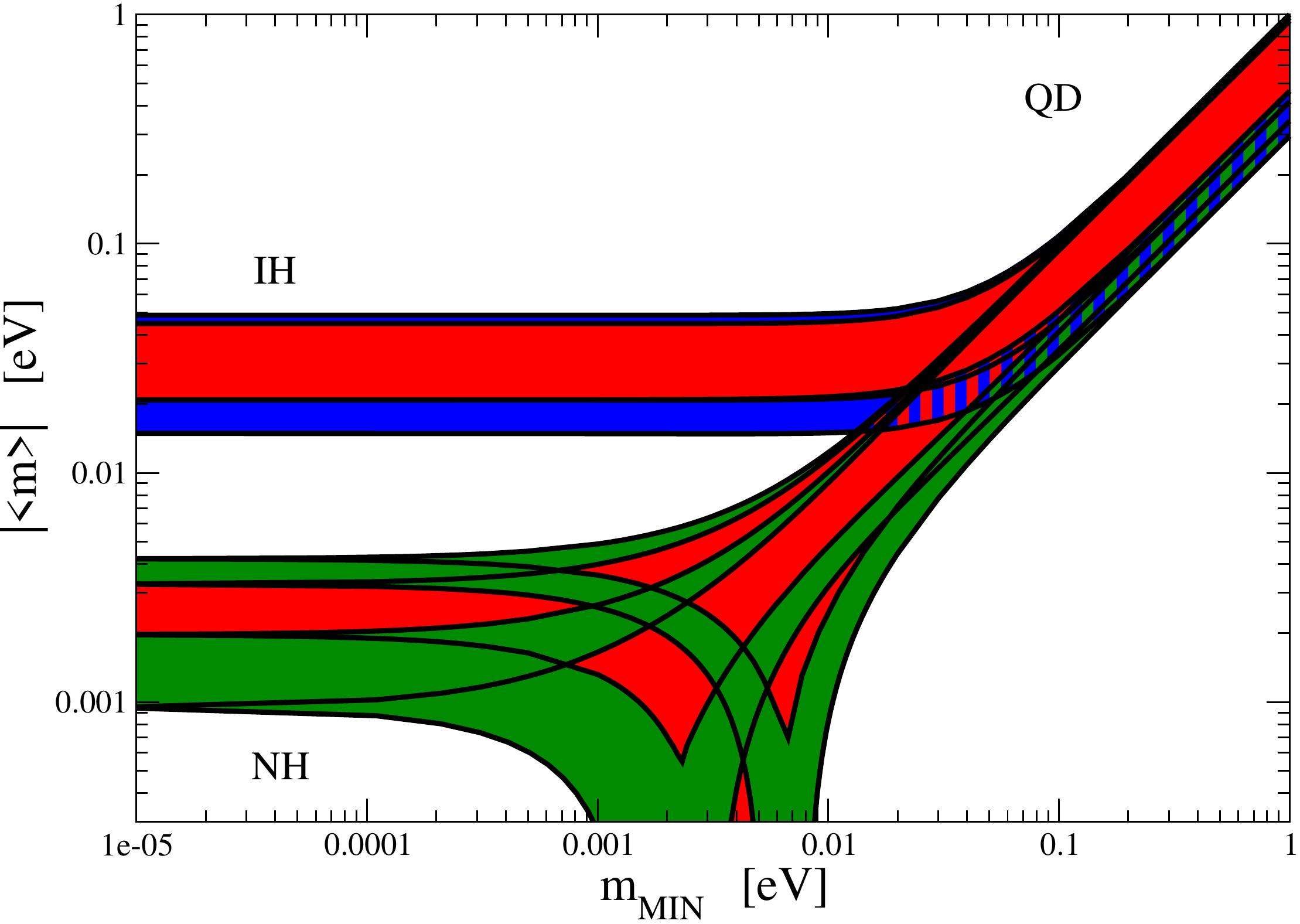}
\caption{
The effective Majorana mass 
$\meff$ (including a 2$\sigma$ uncertainty),
as a function of ${\rm min}(m_j)$ for 
$\sin^2\theta_{13}$ = 0.0236 $\pm 0.0042$ 
\cite{01-An:2012eh} and $\delta =0$. 
The figure is obtained using also 
the best fit values and 1$\sigma$ errors of 
$\Delta m^2_{21}$, $\sin^2\theta_{12}$, and
$|\Delta m^2_{31(32)}|$
given in Table 1.7 in ref. \cite{01-PDG12}. 
The phases $\alpha_{21,31}$
are varied in the interval [0,$\pi$].
The predictions for the NH, IH and QD
spectra are indicated.
The red regions correspond to at least one of
the phases $\alpha_{21,31}$
and $(\alpha_{31} - \alpha_{21})$
having a CP violating value, while the 
blue and green areas correspond to $\alpha_{21,31}$ 
possessing CP conserving values.
(From ref. \cite{01-PDG12}.)
}
\label{Fig1}
\end{figure}
%

   The problem of obtaining 
the allowed values of 
$\meff$ given the constraints on the 
parameters following from $\nu$-oscillation data,
and more generally of the physics potential of 
\betabeta-decay experiments,
was first studied in \cite{01-SPAS94} 
and subsequently in a large number of papers 
\footnote{Extensive list of references on the subject 
is given in \cite{01-WRodej10}.} 
(see, e.g., \cite{01-BPP1,01-PPSchw05,01-PPNH06,01-Other,01-Duerr:2011zd}). 
The results of this analysis are illustrated in Fig. 1. 
The main features of the predictions for 
$\meff$ in the cases of the NH, IH and QD spectra    
are summarised  below.\\
i) NH spectrum:   
\begin{eqnarray}
&&\meff \cong |(\Delta m^2_{21})^{\frac{1}{2}}\, s^2_{12} + 
(\Delta m^2_{31})^{\frac{1}{2}}\, 
s^2_{13}\, e^{-i(\alpha_{21} - \alpha_{31} + 2\delta)}|\,. 
\label{01-meffNH}
\end{eqnarray}
%
Using the $3\sigma$ allowed ranges of the 
relevant neutrino oscillation parameters we get:
\begin{eqnarray}
&& 4.7\times 10^{-4}~{\rm eV}\ltap \meff \ltap 4.8\times 10^{-3}{\rm eV}\,,
~~{\rm NH}\,.
\label{01-meffNH2}
\end{eqnarray}
%
ii) IH spectrum:  
%
\begin{eqnarray}
&& \meff \cong (|\Delta m^2_{32}|)^{\frac{1}{2}}\,
(1-\sin^22\theta_{21}\,\sin^2\frac{\alpha_{21}}{2})^{\frac{1}{2}}\,,\\[0.25cm]
&& (|\Delta m^2_{32}|)^{\frac{1}{2}}\,\cos2\theta_{12}\ltap \meff 
\ltap (|\Delta m^2_{32}|)^{\frac{1}{2}}\,. 
\label{01-meffIH}
\end{eqnarray}
%
Numerically one finds:
\beq 
0.014~{\rm eV} \ltap \meff \ltap 0.050~{\rm eV}\,,~~~{\rm IH}\,,
\label{01-meffIH2}
\eeq
%
the upper and the lower bounds corresponding to the CP-conserving 
values of $\alpha_{21} = 0;~\pi$.\\
iii) QD spectrum: 
\begin{eqnarray}
&& \meff \cong m_0\,(1 - \sin^22\theta_{12}\,
\sin^2\frac{\alpha_{21}}{2})^{\frac{1}{2}}\,,\\[0.25cm]
&& m_0 \gtap \meff \gtap m_0\, \cos2\theta_{12} \gtap 0.028~{\rm eV}\,,
\label{01-meffQD}
\end{eqnarray}
%
with $m_0 \gtap 0.1$ eV, $m_0 < 2.05$ eV \cite{01-MoscowH3b} 
(see also \cite{01-MainzKATRIN}), 
or $m_0\ltap (0.3 - 1.3)$ eV \cite{01-summj}
(see eq. (\ref{01-H3beta}) 
and the discussion following after it).\\

  For the IH (QD) spectrum we have also
\cite{01-BGKP96,01-BPP1}: 
\beq
\sin^2(\frac{\alpha_{21}}{2}) \cong
 \left (1 - \frac{\meff^2}{\tilde{m}^2}\right)\,
\frac{1}{\sin^{2}2\theta_{12}}\,,~~ 
\tilde{m}^2 \equiv |\Delta m^2_{32}|~(m_0^2)\,,~~~{\rm IH~(QD)}\,.
\label{01-alpha21}
\eeq
%
Thus, a measurement of $\meff$ and $m_0$ 
($|\Delta m^2_{32}|$) for QD (IH) spectrum 
can allow to determine $\alpha_{21}$. \\

 The experimental searches for $\betabeta$-decay  
have a long history (see, e.g., \cite{01-bb0nuExp1}).
The most stringent upper limits on $\meff$ were set by the 
IGEX \cite{01-IGEX00}, CUORICINO  \cite{01-CUORI}, 
NEMO3 \cite{01-NEMO3} and EXO-200 \cite{01-EXO12} 
experiments with  $^{76}$Ge,  $^{130}$Te, $^{100}$Mo and 
$^{136}$Xe, respectively 
\footnote{The NEMO3 collaboration has searched for 
$\betabeta$-decay of  $^{82}$Se and other isotopes as well.
}.   
The IGEX collaboration has obtained for the half-life of 
$^{76}$Ge,  $T_{1/2}^{0\nu} > 1.57\times 10^{25}$~yr (90\%~C.L.),
from which the limit 
$\meff < (0.33 - 1.35)$~eV was derived \cite{01-IGEX00}. 
Using the recent more advanced calculations 
of the corresponding nuclear matrix elements (including the relevant 
uncertainties) \cite{01-SFMRS09} 
one finds:  $\meff < (0.22 - 0.35)$~eV.
The NEMO3 and CUORICINO experiments,
designed to reach a sensitivity to $\meff\sim (0.2-0.3)$ eV, 
set the limits: $\meff < (0.61\,- \,1.26)$~eV~\cite{01-NEMO3} and 
$\meff < (0.19 - 0.68)$~eV~\cite{01-CUORI} (90\% C.L.), 
where estimated uncertainties in the NME are accounted for. 
The two upper limits were derived from the
experimental lower limits on the half-lives of $^{100}$Mo and
$^{130}$Te, $T_{1/2}^{0\nu} > 5.8\times 10^{23}$~yr 
(90\%C.L.)~\cite{01-NEMO3} and 
$T_{1/2}^{0\nu} > 3.0\times 10^{24}$~yr
(90\%C.L.)~\cite{01-CUORI}. 
With the NMEs and their uncertainties calculated 
in \cite{01-SFMRS09}, the NEMO3 and CUORICINO upper limits read, 
respectively:  $\meff < (0.50 - 0.96)$~eV and 
$\meff < (0.25 - 0.43)$~eV. 
A very impressive lower limit on the half-life 
of $^{136}$Xe was obtained recently in 
the EXO-200 experiment \cite{01-EXO12}:
$T_{1/2}^{0\nu}(^{136}Xe) > 1.6\times 10^{25}$~yr (90\% C.L.).

The best lower limit on the half-life of $^{76}$Ge,
$T_{1/2}^{0\nu} > 1.9\times 10^{25}$~yr (90\%~C.L.),
was found in the Heidelberg-Moscow $^{76}$Ge 
experiment~\cite{01-HMGe76}. 
It corresponds to the upper limit \cite{01-SFMRS09}
$\meff < (0.20 - 0.35)$~eV.
A positive $\betabeta$-decay signal at $> 3\sigma$,
corresponding to $T_{1/2}^{0\nu} = (0.69 - 4.18)\times
10^{25}$~yr (99.73\%~C.L.) and implying $\meff = (0.1 -
0.9)~{\rm eV}$, is claimed to have been observed in 
\cite{01-KlapdorMPLA}, while a later analysis reports 
evidence for $\betabeta$-decay at 6$\sigma$ 
corresponding to $\meff = 0.32 \pm 0.03$~eV~\cite{01-Klap04}. 

 Most importantly, a large number of projects aim at a
sensitivity to $\meff \sim (0.01 - 0.05)$ eV \cite{01-bb0nuExp2}: CUORE
($^{130}$Te), GERDA ($^{76}$Ge), SuperNEMO, EXO ($^{136}$Xe), MAJORANA
($^{76}$Ge), MOON ($^{100}$Mo), COBRA ($^{116}$Cd), XMASS
($^{136}$Xe), CANDLES ($^{48}$Ca), KamLAND-Zen ($^{136}$Xe), 
SNO+ ($^{150}Nd$), etc.  These experiments, in
particular, will test the positive result claimed 
in \cite{01-Klap04}. 

  The existence of significant 
lower bounds on $\meff$
in the cases of IH and QD spectra 
\cite{01-PPSNO2bb},
which lie either partially (IH spectrum) or completely
(QD spectrum) within the range of sensitivity of 
the next generation of \betabeta-decay experiments,
is one of the most important features of
the predictions of $\meff$. 
These minimal values are given, up to small corrections, 
by $|\Delta m^2_{32}| \cos2\theta_{12}$ and
$m_0 \cos2\theta_{12}$. According to the
combined analysis of the solar and reactor 
neutrino data \cite{01-Fogli:2012XY},
i) the possibility of  $\cos2\theta_{12}$ = 0
is excluded at $\sim 6\sigma$,
ii) the best fit value of 
$\cos2\theta_{12}$ is 
$\cos2\theta_{12} \cong 0.39$, and
iii) at 99.73\% C.L. one has
$\cos{2\theta_{12}} \gtap 0.28$.
The quoted results on $\cos{2\theta_{12}}$
together with the range of possible values of 
$|\Delta m^2_{32}|$ and $m_0$ 
lead to the conclusion about the existence of 
significant and robust lower 
bounds on $\meff$ in the cases of 
IH and QD spectrum. 
At the same time 
one can {\it always} have $\meff \ll 10^{-3}$ eV
in the case of spectrum with 
normal ordering \cite{01-PPW}. As Fig. 1 indicates, 
$\meff$ cannot exceed $\sim 5$ meV
for NH neutrino mass spectrum. 
This implies that $max(\meff)$ in the case of NH 
spectrum is considerably smaller than
$min(\meff)$ for the IH and QD spectrum.
This opens the 
possibility of obtaining 
information about the type of 
$\nu$-mass spectrum from a measurement of 
$\meff \neq 0$
~\cite{01-PPSNO2bb}.
In particular, a positive result in the future 
\betabeta-decay experiments with $\meff > 0.01$ eV
would imply that the 
NH spectrum is strongly disfavored (if not excluded).
For $\Delta m^2_{31(32)} > 0$, 
such a result would mean that the 
neutrino mass spectrum is 
with normal ordering, but is 
not hierarchical. If $\Delta m^2_{31(32)} < 0$, 
the neutrino mass spectrum would be 
either IH or QD. 
Prospective experimental errors 
in  the values of oscillation parameters
in $\meff$ and the sum of neutrino masses,
and the uncertainty in the relevant NME,
can weaken but do not invalidate these results 
\footnote{Encouraging results,
in what regards the problem of 
calculation of the NME, were reported 
at the MEDEX'11 Workshop on Matrix Elements for the 
Double-beta-decay Experiments \cite{01-MEDEX11}.
For the bounds on $\meff$ obtained using the 
current results on the NME see, e.g., \cite{01-LisiFess08}.}
\cite{01-PPRSNO2bb,01-PPSchw05,01-LisiFess08}.

 As Fig. 1 indicates, a measurement
of $\meff \gtap 0.01$ eV would either \cite{01-PPW}
i) determine a relatively narrow 
interval of possible values 
of the lightest neutrino mass ${\rm min}(m_j)$, or 
ii) would establish an upper limit on  ${\rm min}(m_j)$. 
If an upper limit on $\meff$ is experimentally 
obtained below 0.01 eV,
this would lead to a significant upper limit
on  ${\rm min}(m_j)$.

  The possibility of establishing  CP-
violation in the lepton sector 
due to Majorana CPV  
phases has been studied in \cite{01-PPW,01-BargerCP} and in 
much greater detail in \cite{01-PPR1,01-PPSchw05}.
It was found that it is very challenging:
it requires quite accurate measurements 
of $\meff$ (and of $m_0$ for QD spectrum), 
and holds only for a limited range of 
values of the relevant parameters.
More specifically \cite{01-PPR1,01-PPSchw05}, 
establishing at 2$\sigma$
CP-violation associated with
Majorana neutrinos 
in the case of QD spectrum requires 
for $\sin^2\theta_{\odot}$=0.31, in particular, 
a relative experimental 
error on the measured value of 
$\meff$ and $m_0$ smaller than 15\%,
a ``theoretical uncertainty'' 
$F{\small \ltap}$1.5 in the value of
$\meff$ due to an imprecise 
knowledge of the corresponding NME, 
and value of the relevant Majorana
CPV phase $\alpha_{21}$ typically within the ranges 
of ${\small\sim (\pi/4 - 3\pi/4)}$ and
${\small\sim (5\pi/4 - 7\pi/4)}$. 

The knowledge of NME with 
sufficiently small uncertainty 
\footnote{A possible test of 
the NME calculations is suggested in \cite{01-PPW} and 
is discussed in greater detail in \cite{01-NMEBiPet04}.
}
is crucial for obtaining quantitative information on
the $\nu$-mixing parameters from a measurement of
$\betabeta$-decay half-life. The observation of
a $\betabeta$-decay of one nucleus is likely to lead
to the searches and eventually to observation
of the decay of other nuclei.
One can expect that such a progress, in particular,
will help to solve completely the problem
of the sufficiently precise calculation
of the nuclear matrix elements
for the $\betabeta$-decay \cite{01-PPW}.

 If the future $\betabeta$-decay experiments 
show that $\meff < 0.01$ eV, both the IH and
the QD spectrum will be ruled out for massive 
Majorana neutrinos. If in addition it is 
established in neutrino oscillation 
experiments that the neutrino mass spectrum is 
with {\it inverted ordering}, i.e. 
that $\Delta m^2_{31(32)} < 0$,
one would be led to conclude that 
either the massive neutrinos $\nu_j$ 
are Dirac fermions, or that 
$\nu_j$ are Majorana particles
but there are additional contributions to the 
\betabeta-decay amplitude which 
interfere destructively 
with that due to the exchange of 
light massive Majorana neutrinos. 
The case of more than one mechanism 
generating the $\betabeta$-decay was discussed 
recently in, e.g., \cite{01-AMMultiple11,01-ELMultiple11},
where the possibility to identify the mechanisms 
inducing the decay was also analised.
If, however, $\Delta m^2_{31(32)}$ 
is determined to be positive
in neutrino oscillation experiments, 
the upper limit $\meff < 0.01$ eV 
would be perfectly compatible with 
massive Majorana neutrinos
possessing NH mass spectrum, 
or mass spectrum with normal ordering but
partial hierarchy, and the quest for $\meff$ 
would still be open.

\indent  If indeed in the next generation of 
\betabeta-decay experiments it is 
found that $\meff < 0.01$ eV, 
while the neutrino oscillation experiments 
show that $\Delta m^2_{31(32)} > 0$, 
the next frontier in the searches for 
$\betabeta-$decay would most probably 
correspond to values of $\meff \sim 0.001$ eV.
Taking $\meff = 0.001$ eV as a reference value,
the conditions under which $\meff$ in the 
case of neutrino mass spectrum with normal 
ordering would be guaranteed to satisfy 
$\meff \gtap 0.001$ eV, 
were investigate in \cite{01-PPNH06}.
In the analysis performed in \cite{01-PPNH06},
the specific case of 
normal hierarchical neutrino mass spectrum, 
and the general case of spectrum 
with normal ordering, partial hierarchy 
and values of $\theta_{13}$, 
including the value measured 
in the Daya Bay, RENO, 
Double Chooz and T2K experiments, 
eq. (\ref{01-DBayth13}),
were considered.
The ranges of the lightest 
neutrino mass $m_1$ and/or of 
$\sin^2\theta_{13}$, for which 
$\meff \gtap 0.001$ eV were derived 
as well, and the phenomenological 
implications of such scenarios 
were discussed. 

\section{Outlook}


 The last 14 years or so witnessed 
a spectacular experimental progress 
in the studies of the properties of neutrinos.
In this period the existence of neutrino oscillations, 
caused by nonzero neutrino masses and neutrino mixing, 
was established and the parameters which drive the oscillations, 
were determined with a relatively high precision.
In spite of these remarkable achievements one has to admit that 
we are still completely ignorant about some of the fundamental 
aspects of neutrino mixing: the nature - Dirac or Majorana, 
of massive neutrinos, the type of spectrum the neutrino masses 
obey, the absolute scale of neutrino masses, the status of 
CP symmetry in the lepton sector.
Finding out these aspects and understanding 
the origins of the neutrino masses and mixing 
and the patterns they and possibly leptonic CP 
violation exhibit, requires an extensive and challenging 
program of research.
The main goals of such a research program
include:\\
\begin{itemize}
\item Determining
the nature - Dirac or Majorana, of massive neutrinos $\nu_j$.
This is of fundamental importance for making progress in
our understanding of the origin of
neutrino masses and mixing and
of the symmetries governing the lepton sector
of particle interactions.
\item{Determination of the
sign of $\Delta m^2_{31(32)}$
$(\Delta m^2_{31})$ and of
the type of 
neutrino mass spectrum.}
\item{Determining or
obtaining significant constraints
on the absolute neutrino mass scale.}
\item{
  Determining the status of CP symmetry
              in the lepton sector.}
\item{Understanding at a fundamental level
the mechanism giving rise to
neutrino masses and mixing and to
$L_l-$non-conservation.
This includes
understanding the origin of the
patterns of 
neutrino mixing and neutrino masses, 
suggested by the data.
Are the observed patterns of
$\nu$-mixing and of $\Delta m^2_{21,31}$
related to the existence of a new
fundamental symmetry of particle interactions?
Is there any relation between quark
mixing and neutrino (lepton) mixing?
What is the physical origin
of CP violation phases in
the neutrino mixing matrix $U$?
Is there any relation (correlation)
between the (values of) CP violation
phases and mixing angles in  $U$?
Progress in the theory of
neutrino mixing might also lead
to a better understanding of the
mechanism of generation of baryon
asymmetry of the Universe.}
\end{itemize}
 
 The successful realization
of this research program
would be a formidable task and would require many years. 
It already began 
with the high precision measurement
of  $\theta_{13}$ in the Daya Bay and RENO 
experiments, which showed that  $\sin^22\theta_{13}$
has a relatively large value, eq. (\ref{01-DBayth13}).
The Double Chooz and T2K experiments also 
found values of $\sin^22\theta_{13}$, 
which are different from zero respectively at 
$3.1\sigma$ and $3.2\sigma$ and are compatible 
with those obtained in the Daya Bay and RENO 
experiments.
These results on $\theta_{13}$ have far reaching 
implications. As we have already mentioned or 
discussed, the measured relatively large 
value of $\theta_{13}$ opens up the 
possibilities, in particular,

i) for searching for CP violation effects in 
neutrino oscillation experiments with 
high intensity accelerator 
neutrino beams, like T2K and NO$\nu$A 
\footnote{The sensitivities of T2K and NO$\nu$A on CP violation 
in neutrino oscillations are discussed, e.g., 
in \cite{01-JBerna2010}.
} 
\cite{01-NOvAproposal}, 

ii) for determining the sign of 
$\Delta m^2_{32}$, and thus the type 
of neutrino mass spectrum, 
in neutrino oscillation experiments
with sufficiently long baselines 
(see, e.g., \cite{01-Future,01-ReactNuHiera}).

 A value of $\sin\theta_{13} \gsim 0.09$ 
is a necessary condition for a successful ``flavoured''
leptogenesis with hierarchical heavy Majorana neutrinos
when the CP violation required for 
the generation of the matter-antimatter asymmetry of the
Universe is provided entirely by the Dirac CP violating
phase in the neutrino mixing matrix \cite{01-PPRio106}.

  With the measurement of $\theta_{13}$, 
the first steps on the long ``road''
leading to a comprehensive understanding
of the patterns of neutrino masses and
mixing, of their origin and 
implications, were made. 
The future of neutrino physics is bright.

\section{Acknowledgements.}
This research was supported in part by the INFN program on
``Astroparticle Physics'', by the Italian MIUR program on
``Neutrinos, Dark Matter and  Dark Energy in the Era of LHC''
by the World Premier International Research Center 
Initiative (WPI Initiative), MEXT, Japan, and 
by the European Union FP7-ITN INVISIBLES 
(Marie Curie Action, PITAN-GA-2011-289442).


\end{document}